\documentclass[aip,pop,groupedaddress,preprint]{revtex4-1}
\usepackage{physics}
\usepackage{amsmath}
\usepackage{graphicx}
\usepackage{subfigure}
\usepackage{hyperref}
\begin{document}
\title{Stationary states of polytropic plasmas} 
\author{Ran Guo}
\thanks{Author to whom correspondence should be addressed}
\email{rguo@cauc.edu.cn }
\affiliation{Department of Physics, College of Science, Civil Aviation University of China, Tianjin 300300, China}
\begin{abstract}
   In this paper, we prove that the kappa distribution is the stationary solution of the Vlasov-Poisson system in an inhomogeneous plasma under the polytropic equation of state and an assumption restricting the local velocity distribution to a specific mathematical form.
   The profiles of density, temperature, and electric potential are obtained theoretically.
   The kappa index can be determined if the initial state is known.
   In order to verify the theory, particle-in-cell simulations are made and the results show excellent agreement with the theoretical predictions for density, temperature, and velocity distributions of electrons.
   It is shown that the electron velocity distribution of spatially inhomogeneous plasma evolves from an initial Maxwellian to the final kappa distribution.
   It is also found that the value of kappa index in the final stationary state depends on the initial state of plasma.
\end{abstract}
\maketitle

\section{Introduction}
Space plasmas are typical collisionless many-body systems far away from thermal equilibrium. 
Due to the long-range electromagnetic interactions, the suprathermal particles can be generated in lots of situations and the velocity distribution functions of those plasmas are usually different from Maxwellian distribution. 
A more suitable model to describe those suprathermal particles is the kappa distribution. As a generalization of Maxwellian distribution, the kappa distribution has been used to model many different space plasmas such as solar corona, \cite{Vocks2008AA,Cranmer2014AJL} solar wind, \cite{Maksimovic1997AA,Leubner2004AJ,Pierrard2016SP} planetary magnetosphere, \cite{Collier1995GRL,Schippers2008JGRSP,Dialynas2009JGRSP} Io plasma torus, \cite{STEFFL2004I} outer heliosphere, \cite{Decker2005S,Kollmann2019AJ} etc. 
More observational evidence can be found in the Refs. \cite{Pierrard2010SP,Livadiotis2015JGRSP}.
In addition, it is an interesting topic that the kappa-distributed plasma often exhibit remarkably different behaviors compared to the plasma in thermal equilibrium. \cite{Treumann2004PP,Du2004PLA,Livadiotis2009JGRSP,Liu2009PP,Liu2009PPa,Lazar2010MNRAS,Gougam2011PP,Vinas2015JGRSP,Livadiotis2019PP,Livadiotis2019AJ} 
Although this model is very successful, the physical mechanism behind it is still under discussion. 
Some works suggest that the formation of suprathermal particles is due to the interactions between waves and particles. \cite{Hasegawa1985PRL,Ma1998GRL,Vocks2003AJ,Bian2014AJ,Shizgal2018PRE}
Besides, some other investigations explain the generation of kappa distribution from the view of nonextensive statistical mechanics. \cite{Leubner2004PP,Leubner2005AJ,Livadiotis2009JGRSP}

In most space plasmas, the density and temperature vary considerably so that the plasmas need to be recognized as inhomogeneous. 
The changes in density or temperature cause the non-uniform plasmas to present some attractive physical phenomenons such as drift waves and the corresponding instabilities.
Many works reveal that there is a specific relation between inhomogeneity and the kappa distribution.
Some of these works \cite{Scudder1992,MeyerVernet1995,Moncuquet2002,Pierrard2010SP,Livadiotis2016} suggest that a kappa-distributed plasma leads to a polytropic equation of state (EOS) between the density and temperature.
Du's work \cite{Du2004PLA} proves that the inhomogeneous temperature can exist in a kappa-distributed stationary plasma but cannot exist in a Maxwellian one.
Furthermore, it is also pointed out by Livadiotis \cite{Livadiotis2019AJ} that the polytropic EOS can lead to the kappa distribution by using the hydrodynamic method.

The present work is different from these previous studies.
In this work, we investigate the stationary state of an inhomogeneous isotropic plasma by both theoretical derivations and particle-in-cell(PIC) simulations. 
The stationary electron kappa distribution is derived by solving the kinetic Vlasov-Poisson equations under two assumptions: (i) the electron density and temperature follow the polytropic EOS; (ii) the electron local distribution has a specific form introduced in Section \ref{subsec:model}.
The expressions for the profiles of density, temperature, and electric potential are also obtained.
We then verify the theoretical results by PIC simulations which show excellent agreement.
The most interesting results in this work are that the plasma initially following Maxwellian distribution with inhomogeneous density will evolve to a kappa-distributed plasma in which both density and temperature are non-uniform.
At the final stationary state, the kappa index of distribution function can be determined theoretically from the conservations of the system, which means there are no fitting parameters in our work.

The paper is organized as follows.
In section 2, we introduce our model and derive the solution of the Vlasov-Poisson equations with a polytropic EOS.
In section 3, PIC simulations are performed to verify the theory.
In section 4, the influence of different initial states is discussed. 
At last, we make the conclusions and discussions in section 5.

\section{Theoretical Model} 
\subsection{Model and Definitions}
\label{subsec:model}
We consider an isotropic electrostatic plasma, in which both ions and electrons are spatially inhomogeneous.
The system is assumed to be neutral globally, but possibly non-neutral at local.
For simplicity, we set the ions as a stationary background in the present study.
When the system relaxes to a steady state, the electron distribution $f_e(\vb{r},\vb{v})$ should satisfy the stationary Vlasov-Poisson equations,
\begin{equation}
    \vb{v} \cdot \nabla f_e(\vb{r},\vb{v}) + \frac{e\nabla \varphi(\vb{r})}{m} \cdot  \pdv{f_e(\vb{r},\vb{v})}{\vb{v}} = 0,
    \label{eq:V}
\end{equation}
\begin{equation}
    \nabla^2 \varphi(\vb{r}) = - \frac{e}{\varepsilon_0}[n_i(\vb{r})-n_e(\vb{r})],
    \label{eq:P}
\end{equation}
where $m$ is the electron mass, $e$ is the elementary charge, and $\varphi(\vb{r})$ is the electrostatic potential.
$n_e(\vb{r})$ and $n_i(\vb{r})$ are the number density of electrons and ions respectively. 
Generally speaking, we can always write the stationary distribution function as a product of two terms, i.e. $f_e ( \vb{r}, \vb{v})  =  n_e( \vb{r}) \hat{f}_{e}( \vb{r}, \vb{v})$.
The first term is the density distribution,
\begin{equation}
    n_e( \vb{r}) = \int f_e( \vb{r}, \vb{v}) \dd{\vb{v}}.
    \label{def:ne}
\end{equation}
The second term $\hat{f}_{e}(\vb{r}, \vb{v})$ describes a local velocity distribution at the position $\vb{r}$, which is defined as
\begin{equation}
    \hat{f}_{e}(\vb{r},\vb{v}) = \frac{f_e(\vb{r},\vb{v})}{n_e(\vb{r})}.
    \label{def:fe}
\end{equation}
We will assume that the local velocity distribution $\hat{f}_e(\vb{r},\vb{v})$ is of the form,
\begin{equation}
    \hat{f}_{e}(\vb{r},\vb{v}) = A[T(\vb{r})] g(W^*),
    \label{assump:argu}
\end{equation}
where $T(\vb{r})$ is the local temperature,
\begin{equation}
    \frac{d}{2} k_B T(\vb{r}) = \frac{1}{2}  m \int v^2 \hat{f}_e(\vb{r},\vb{v}) \dd{\vb{v}},
    \label{def:T}
\end{equation}
with the spatial dimension $d = 1,2,3$, 
$W^*$ is the reduced kinetic energy,
\begin{equation}
    W^* = \frac{mv^2}{2k_BT(\vb{r})},
    \label{def:wk}
\end{equation}
$A[T(\vb{r})]$ is the normalization determined by the integral,
\begin{equation}
   A[T(\vb{r})] = \left[ \int g (W^*) \dd{\vb{v}} \right]^{-1}, 
\end{equation}
to ensure $\int \hat{f}_e \dd{\vb{v}} =1$.
$g(W^*)$ is constrained by,
\begin{equation}
   \frac{\int W^* g(W^*) \dd{\vb{v}}}{\int g(W^*) \dd{\vb{v}}} = \frac{d}{2}, 
   \label{constraint:g}
\end{equation}
according to the definition of the local temperature \eqref{def:T}.
The assumption \eqref{assump:argu} with Eqs. \eqref{def:wk}-\eqref{constraint:g} is valid for most distribution functions observed in space plasmas, such as Maxwellian distribution, kappa distribution, Cairns distribution, \cite{Cairns1995GRL} etc.
This assumption also suggests a static stationary solution which means the plasma has no flow velocity, i.e., 
\begin{equation}
    \int \vb{v} f_e(\vb{r},\vb{v}) \dd{\vb{v}} = 0,
\end{equation}
because $f_e(\vb{r},\vb{v})$ is an even function of $\vb{v}$ due to Eqs. \eqref{assump:argu} and \eqref{def:wk}.

Besides the hypothesis on the mathematical form of the local velocity distribution, we also need to assume the relationship between the density and temperature of electrons in the stationary state.
Generally speaking, the exact correlation between density and temperature in an inhomogeneous collisionless plasma is still a complex mystery.
However, it is shown that the polytropic model makes a good job of describing this correlation by a number of observations and experiments in different plasmas. \cite{Sittler1980,Belmont1992,MeyerVernet1995,Erents2000,Liu2006,Nicolaou2014,Pang2016}
Therefore, we assume that the polytropic EOS holds in the stationary state,
\begin{equation}
    \frac{T(\vb{r})}{T_0} = \left[ \frac{n_e(\vb{r})}{n_0} \right]^{\gamma-1},
    \label{eq:poly-rela}
\end{equation}
where $\gamma$ is the polytropic index, $n_{0}=n_e(\vb{r}_0)$ and $T_0=T(\vb{r}_0)$ are the density and temperature at an arbitrary reference point $\vb{r}_0$. 
The index $\gamma$ varies in a large interval and stands for different thermodynamic relations, such as $\gamma=1$ for an isothermal case, $\gamma=5/3$ for an adiabatic relation and the exponent $\gamma-1$ can also take negative values in some cases \cite{MeyerVernet1995,Riley2001,Pang2016}.
Hereinafter we choose $n_{0}$ as the average number density, then $T_0$ turns out to be the corresponding temperature.
This choice does not affect other results obviously.

\subsection{Velocity Distribution}
With the above definitions and assumptions, we start to solve the local velocity distribution from the stationary Vlasov equation.
Substituting Eqs. \eqref{def:fe} and \eqref{assump:argu} into Eq. \eqref{eq:V}, we obtain
\begin{equation}
   \vb{v} \cdot ( A g \nabla n_e + n_e A \nabla g + n_e g \nabla A ) + n_e A \frac{e\nabla \varphi}{m}  \cdot \pdv[]{g}{\vb{v}}=0.
   \label{dev1}
\end{equation}
Dividing the above Eq. \eqref{dev1} by $f_e(\vb{r},\vb{v})$, one can rewrite the Vlasov equation as
\begin{equation}
   \vb{v} \cdot \left( \frac{\nabla n_e}{n_e} + \frac{\nabla g}{g} + \frac{\nabla A}{A} \right) + \frac{e\nabla \varphi}{m} \frac{1}{g} \pdv[]{g}{\vb{v}}  =0.
   \label{dev2}
\end{equation}
With the aid of Eq. \eqref{assump:argu}, the partial derivative of $g$ can be calculated by the chain rule,
\begin{equation}
    \pdv{g}{\vb{v}} = \frac{m\vb{v}}{k_B T} \dv[]{g}{{W^*}} ,
    \label{eq:drhodv}
\end{equation}
\begin{equation}
    \nabla g = - \frac{mv^2}{2k_BT^2}\dv{g}{W^*}\nabla T.
    \label{eq:drhodr}
\end{equation}
Similarly, the gradient of the normalization factor $A$ is,
\begin{equation}
    \nabla A = \dv[]{A}{T} \nabla T.
    \label{eq:dAdr}
\end{equation}
Substituting Eqs. \eqref{eq:drhodv}, \eqref{eq:drhodr} and \eqref{eq:dAdr} into \eqref{dev2}, we derive,
\begin{equation}
   \vb{v} \cdot \left(
    \nabla \ln n_e - \frac{mv^2}{2k_B T} \dv{\ln g}{{W^*}} \nabla \ln T + T\dv{\ln A}{T} \nabla \ln T+ \frac{e \nabla \varphi}{k_B T}\dv[]{\ln g}{{W^*}}
       \right) =0.
   \label{dev3}
\end{equation}
It is worth noting that Eq. \eqref{dev3} is essentially the stationary Vlasov equation \eqref{eq:V} associated with our assumptions Eqs. \eqref{def:fe} and \eqref{assump:argu}. 
Because $\vb{v}$ is an arbitrary vector, the terms in the bracket should vanish,
\begin{equation}
    \nabla \ln n_e - \frac{mv^2}{2k_B T} \dv{\ln g}{{W^*}} \nabla \ln T + T\dv{\ln A}{T} \nabla \ln T+ \frac{e \nabla \varphi}{k_B T}\dv[]{\ln g}{{W^*}}=0.
    \label{dev5}
\end{equation}
Due to the electron thermal motion, there exists a pressure gradient produced by the inhomogeneous density and temperature in the plasma. 
In the stationary state, the gradient of pressure is balanced by the electric field,
\begin{equation}
    n_e e \nabla \varphi = \nabla p,
    \label{eq:hydro-equi}
\end{equation}
where the pressure is considered as a scalar in an isotropic plasma and defined by $p = p(\vb{r}) = \frac{1}{d} mn_e(\vb{r}) \int v^2 \hat{f}_e(\vb{r},\vb{v}) \dd{\vb{v}}$ as common. 
Combining this pressure definition with the temperature definition \eqref{def:T}, one finds that the ideal gas equation of state $p = n_e k_B T$ holds.
Substituting the pressure balance condition \eqref{eq:hydro-equi} into Eq. \eqref{dev5} and utilizing the ideal gas equation of state $p = n_e k_B T$, we obtain,
\begin{equation}
    \nabla \ln n_e - \frac{mv^2}{2k_B T} \dv{\ln g}{{W^*}} \nabla \ln T + T\dv{\ln A}{T} \nabla \ln T+ (\nabla \ln n_e + \nabla \ln T) \dv[]{\ln g}{{W^*}}=0.
    \label{dev6}
\end{equation}
By using the definition of normalization, we can prove an identity
\begin{equation}
    T\dv{\ln A}{T} = -\frac{d}{2}. 
    \label{eq:identity}
\end{equation}
The proof of the identity is provided in the Appendix \ref{ap:equality} in detail.
So the above equation \eqref{dev6} can be solved if the relationship between density $n_e(\vb{r})$ and temperature $T(\vb{r})$ is known.
As we have introduced in Section \ref{subsec:model}, the polytropic EOS \eqref{eq:poly-rela} is assumed to be valid.
After taking the logarithm on both sides of Eq. \eqref{eq:poly-rela} and calculating the derivative, we have
\begin{equation}
    \nabla \ln T = (\gamma-1) \nabla \ln n_e = \frac{1}{-\kappa_0-1} \nabla \ln n_e,
    \label{eq:poly-rela2}
\end{equation}
where the symbol
\begin{equation}
    \kappa_0 = -1 - \frac{1}{\gamma-1},
    \label{eq:kappa-poly}
\end{equation}
is denoted in order to give a familiar mathematical form of kappa distribution.
Substituting Eqs. \eqref{eq:identity} and \eqref{eq:poly-rela2} into Eq. \eqref{dev6}, we find the equation of $g(W^*)$,
\begin{equation}
    \dv[]{\ln g}{{W^*}} = \frac{-\kappa_0-1-\frac{d}{2}}{\kappa_0} \frac{1}{1 + \frac{W^*}{\kappa_0} }.
    \label{eq:rho}
\end{equation}
The local velocity distribution of electrons can be solved by integrating Eq. \eqref{eq:rho},
\begin{equation}
    g(W^*) = C \left( 1+ \frac{W^*}{\kappa_0} \right)^{-\kappa_0 -1 - \frac{d}{2} }= C \left[ 1+ \frac{mv^2}{\kappa_0 2 k_B T(\vb{r})} \right]^{-\kappa_0 -1 - \frac{d}{2} },
    \label{sol:rho}
\end{equation}
with an integral constant $C$.
The normalization factor can be calculated by integrating Eq. \eqref{sol:rho},
\begin{equation}
    A[T(\vb{r})] = \left[ \frac{m}{2 \pi k_B T(\vb{r})\kappa_0}  \right]^{ \frac{d}{2}  } \frac{\Gamma(\kappa_0+1+ \frac{d}{2} )}{\Gamma(\kappa_0+1)} C^{-1}.
    \label{sol:norm-rho}
\end{equation}
Therefore, the total electron distribution $f_e(\vb{r},\vb{v})$ is,
\begin{equation}
    f_{e} (\vb{r},\vb{v}) = n_e(\vb{r}) \left[ \frac{m}{2 \pi k_B T(\vb{r}) \kappa_0} \right]^ \frac{d}{2}  \frac{\Gamma(\kappa_0+1+ \frac{d}{2})}{\Gamma(\kappa_0+1)}
    \left[ 1+ \frac{1}{\kappa_0} \frac{m v^2}{2k_BT(\vb{r})} \right ]^{ -\kappa_0 - 1 - \frac{d}{2} },
    \label{fun:kappaPDF}
\end{equation}
which exactly agrees with the well-known kappa distribution in Refs. \cite{Livadiotis2015JGRSP,Livadiotis2011AJ,Livadiotis2016}. 
Here $\kappa_0$ is the kappa index that is independent of space or velocity dimensions. \cite{Livadiotis2011AJ}
This solution is consistent with the definition of the local temperature \eqref{def:T}.
The detailed derivation can be found in Appendix \ref{ap:consistency}. 
It is worth noting that there is another form of the kappa velocity distribution commonly used in other works, \cite{Summers1991,Maksimovic1997AA,MeyerVernet2001,Moncuquet2002} 
\begin{equation}
    f_\kappa(\vb{v}) = n \left[\frac{m}{2 \pi k_B T (\kappa - \frac{3}{2})} \right]^{\frac{3}{2}}\frac{\Gamma(\kappa+1)}{\Gamma(\kappa-\frac{1}{2})} \left(
    1 + \frac{1}{\kappa -\frac{3}{2}} \frac{m v^2}{2 k_B T}
    \right)^{-\kappa-1}.
    \label{fun:kappaPDF-1}
\end{equation}
With $\kappa = \kappa_0 + d/2$ and $d=3$, the two kappa distributions \eqref{fun:kappaPDF} and \eqref{fun:kappaPDF-1} are the same.
Therefore, any results derived from the former kappa distribution can be transformed into those from the latter kappa distribution.
In addition, with $\kappa = \kappa_0 + d/2$ and $d=3$, the relationship between the polytrope and the kappa index \eqref{eq:kappa-poly} becomes,
\begin{equation}
    \kappa = \frac{1}{2} - \frac{1}{\gamma-1},
\end{equation}
which also concurs with other Refs. \cite{MeyerVernet1995,Moncuquet2002,MeyerVernet2001}.

It is a remarkable fact that the electron distribution \eqref{fun:kappaPDF} is derived from the Vlasov equation \eqref{eq:V} with the pressure balance \eqref{eq:hydro-equi}, the assumption on the form of the local velocity distribution \eqref{assump:argu}, and the polytropic EOS \eqref{eq:poly-rela2}. 
It is worth to note that the pressure balance is not an extra assumption but a consequence of stationarity, isotropy, and zero-flow. 
Up to this point, only the velocity distribution is solved explicitly.
The electron density $n_e(\vb{r})$ is still unsolved due to the absence of the Poisson equation and the ion density $n_i(\vb{r})$ in the above derivations \eqref{dev1}-\eqref{fun:kappaPDF}.
Therefore, we will study the electron density $n_e(\vb{r})$ through the Poisson equation with the ion density $n_i(\vb{r})$ in the next subsection to ensure the polytropic EOS is imposed in a self-consistent manner.

\subsection{Spatial Distribution}
After working out the velocity distribution, we continue to solve the spatial distribution.
For convenience, we denote $n^*_e(\vb{r}) = n_e(\vb{r})/n_0 $ and $ n^*_i(\vb{r}) = n_i(\vb{r}) /n_0$ where $n^*_e(\vb{r})$ and $n^*_i(\vb{r})$ represent the relative electron and ion density. 
$n_0$ is the average number density for both electrons and ions.
Collecting the Poisson equation \eqref{eq:P}, the pressure balance \eqref{eq:hydro-equi}, and the equation of state $p = n_e k_B T$, we can obtain
\begin{equation}
    \nabla \cdot \left[ \frac{\nabla (n^*_ek_BT)}{n^*_e e} \right] = - \frac{en_0}{\varepsilon_0} (n^*_i - n^*_e).
    \label{eq:ns-dev}
\end{equation}
Substituting the polytropic EOS \eqref{eq:poly-rela} into Eq. \eqref{eq:ns-dev}, one finds the equation of relative electron density,
\begin{equation}
    \nabla^2 n^*_e(\vb{r}) - \frac{\kappa_0+2}{\kappa_0+1}\frac{[\nabla n^*_e(\vb{r})]^2}{n^*_e(\vb{r})} - \frac{\kappa_0+1}{\kappa_0} \frac{n_{0} e^2}{\varepsilon_0 k_B T_0} n^*_e(\vb{r})^{\frac{1}{\kappa_0+1}+1} [n^*_e(\vb{r})-n^*_i(\vb{r})] =0. 
    \label{eq:u}
\end{equation}
If the parameters $n_0$, $\kappa_0$, and $T_0$ are known, and the relative ion density $n^*_i(\vb{r})$ is given, Eq. \eqref{eq:u} is solvable through the numerical method with some certain boundary conditions.

However, without solving the Eq. \eqref{eq:u}, we find an important property of the solution. 
If both the relative electron density $n^*_e(\vb{r})$ and ion density $n^*_i(\vb{r})$ are periodic and smooth functions, we can prove that a necessary condition of inhomogeneous electron density is the non-uniform ion density.
The proof is as follows.
For a periodic and smooth $n^*_e(\vb{r})$, the global maximum of $n^*_e(\vb{r})$ must also be a local maximum.
Suppose that $n^*_e(\vb{r})$ achieves its global maximum at $\vb{r} = \vb{r}_{max}$.
The gradient of the electron density at this extreme point is zero, \cite{Marsden2012}
\begin{equation}
    \nabla n^*_e(\vb{r}_{max}) = 0.
    \label{eq:dv1-extre}
\end{equation}
We use $H(\vb{r})$ to represent the Hessian matrix of $n^*_e$.
At the relative maximum $\vb{r}_{max}$, the Hessian matrix $H(\vb{r}_{max})$ is negative semidefinite, and its trace is nonpositive, namely $\tr H(\vb{r}_{max})  \leq 0$. \cite{Hogben2013}
In the Cartesian coordinate system, it is obvious that $\tr H = \nabla^2 n^*_e$, therefore the Laplacian of electron density satisfies,
\begin{equation}
    \nabla^2 n^*_e(\vb{r}_{max}) \leq 0.
    \label{eq:dv2-extre}
\end{equation}
Substituting Eqs. \eqref{eq:dv1-extre} and \eqref{eq:dv2-extre} into Eq. \eqref{eq:u}, we obtain
\begin{equation}
    n^*_e(\vb{r}_{max}) \leq n^*_i(\vb{r}_{max}),
    \label{eq:neni-neq}
\end{equation}
by using the inequalities $n_0 e^2/(\varepsilon_0 k_B T_0) > 0$, $\kappa_0>0$ , and $n^*_e(\vb{r}_{max}) > 0$.
The inequality of the kappa index $\kappa_0 >0$, which is equivalent to $\kappa > 3/2$ in another form of the kappa distribution \eqref{fun:kappaPDF-1}, is required to ensure a finite temperature. \cite {MeyerVernet2001,Livadiotis2010a}
The global maximum $n^*_e(\vb{r}_{max})$ cannot equal zero, otherwise $n^*_e(\vb{r}) = 0$ for all $\vb{r}$ because $n^*_e(\vb{r})$ is nonnegative.
According to the above inequality \eqref{eq:neni-neq}, if the ion density is a constant, i.e. $n^*_i(\vb{r})=n^*_i$, then $n^*_e(\vb{r}_{max}) \leq n^*_i$. 
In addition, due to the global neutrality, we have $\int [n^*_e(\vb{r})-n^*_i] \dd{\vb{r}} = 0$.
Combining the above two factors, we find that the only possible solution is $n^*_e(\vb{r}) = n^*_i$,
which indicates that the electron density $n^*_e$ must be uniform in a uniform ion background.
Therefore, the inhomogeneous electron density can only exist in a non-uniform ion background.
It is a necessary condition of the inhomogeneity of the electron density.
This conclusion is very important, because a uniform electron density, resulting in a uniform temperature in terms of the polytropic EOS \eqref{eq:poly-rela}, is a trivial solution in this work. 

After solving $n^*_e(\vb{r})$ from Eq. \eqref{eq:u} with appropriate boundary condition, we can calculate the temperature in terms of the electron density from the polytropic equation \eqref{eq:poly-rela}
\begin{equation}
    T(\vb{r}) = T_0 n^*_e(\vb{r})^{- \frac{1}{\kappa_0+1} },
    \label{fn:temper}
\end{equation}
and then the electrostatic potential from the pressure balance \eqref{eq:hydro-equi}
\begin{equation}
    \varphi(\vb{r}) = \varphi_0 - \kappa_0 \frac{k_B T_0}{e} n^*_e(\vb{r})^{- \frac{1}{\kappa_0+1} },
    \label{fn:phi}
\end{equation}
where $\varphi_0$ is an integral constant and can be determined by the zero potential selection. Finally, the electron distribution function is obtained, by replacing the temperature $T(\vb{r})$ in Eq. \eqref{fn:temper},
\begin{equation}
    f_e (\vb{r},\vb{v}) = n_0 n^*_e(\vb{r}) \left[ \frac{m n^*_e(\vb{r})^{ \frac{1}{\kappa_0+1}  }}{2\pi k_B T_0 \kappa_0}  \right]^{ \frac{d}{2} } \frac{\Gamma(\kappa_0+1+\frac{d}{2})}{\Gamma(\kappa_0+1)} \left[ 1+ \frac{1}{\kappa_0}  \frac{mv^2}{2k_BT_0} n^*_e(\vb{r})^{ \frac{1}{\kappa_0+1}  } \right]^{-\kappa_0-1- \frac{d}{2} }.
    \label{eq:pdf}
\end{equation}
From the above derivations of the kappa distribution, we find that the polytropic EOS \eqref{eq:poly-rela}plays a significant role.
Generally speaking, the polytropic EOS describes the relationship between the macroscopic quantities, while the distribution function describes the microscopic state.
They are two different equations.
However, this study indicates that there is a strong relationship between the polytropic EOS and the kappa distribution.
On the one hand, the kappa index $\kappa_0$ is directly related to the polytropic index $\gamma$ through Eq. \eqref{eq:kappa-poly}.
If we take the limitation $\gamma \rightarrow 1$, which is equivalent to $\kappa_0 \rightarrow \infty$, 
the electron distribution \eqref{eq:pdf} recovers the Maxwellian one, and the temperature profile \eqref{fn:temper} becomes uniform.
On the other hand, the kappa distribution \eqref{eq:pdf} is a sufficient condition of the polytropic EOS \eqref{eq:poly-rela}, \cite{Scudder1992,MeyerVernet1995,Moncuquet2002,Pierrard2010SP,Livadiotis2016} while the polytropic EOS is only \emph{a necessary condition} of the kappa distribution.
The sufficient conditions to derive the kappa distribution should be the stationary condition, polytropic EOS \eqref{eq:poly-rela}, inhomogeneity, and assumption \eqref{assump:argu}.
For instance, if we remove the condition of inhomogeneity, i.e., considering a uniform isothermal plasma in which $n_e=n_i=n_0$ and $T=T_0$, then the polytropic EOS becomes a trivial but valid equation $1=1^{\gamma -1}$.
In this case, any function of energy could be the stationary distribution of the Vlasov-Poisson equations.

So far, we have proved under the polytropic EOS \eqref{eq:poly-rela} and the assumption of the local velocity distribution \eqref{assump:argu}, 
the stationary state of the electrons in an inhomogeneous iostropic plasma obeys the kappa distribution \eqref{eq:pdf}.
The electron inhomogeneity, including the density and the temperature, can only exist in a non-uniform ion background.
The electron distribution \eqref{eq:pdf} in which $n^*_e(\vb{r})$ satisfies Eq. \eqref{eq:u} can be regarded as a solution of the coupled Vlasov-Poisson equations \eqref{eq:V} and \eqref{eq:P}.

\subsection{Determination of The Parameters}
In principle, if the initial state of the plasma is known, the final stationary distribution can be determined completely,
including the parameters $\kappa_0$, $T_0$.
In our model, these two quantities $\kappa_0$ and $T_0$ can be derived from the conserved quantities.
As we know, in a collisionless plasma system governed by the Vlasov equation, the total energy $E$
\begin{equation}
    E = \frac{1}{2} \int n_e(\vb{r}) k_B T(\vb{r}) \dd{\vb{r}} + \frac{1}{2} \int [n_i(\vb{r}) - n_e(\vb{r})] e \varphi(\vb{r}) \dd{\vb{r}},
    \label{fn:energy}
\end{equation}
and the Boltzmann-Gibbs entropy $S$
\begin{equation}
    S = -k_B \iint f_e(\vb{r},\vb{v}) \ln f_e(\vb{r},\vb{v}) \dd{\vb{v}} \dd{\vb{r}}
    \label{fn:entropy}
\end{equation}
are the conservations.
In Eqs. \eqref{fn:energy} and \eqref{fn:entropy}, the number density of electrons $n_e(\vb{r})$, the temperature $T(\vb{r})$, and the distribution function $f_e(\vb{r},\vb{v})$ can all be expressed in the form of $n^*_e(\vb{r})$ with two parameters $\kappa_0$ and $T_0$. 
Therefore, we have three Eqs. \eqref{eq:u}, \eqref{fn:energy} and \eqref{fn:entropy} for the three unknown quantities $\kappa_0$, $T_0$ and $n^*_e(\vb{r})$.
In these three equations, the other quantities can be obtained from the determined initial distribution, including the values of energy and entropy, the averaged density $n_0$, and the relative ion density $n^*_i(\vb{r})$.
Solving these three equations simultaneously, we can derive $\kappa_0$, $T_0$ and $n^*_e(\vb{r})$ finally.  
Substituting the results of $\kappa_0$, $T_0$ and $n^*_e(\vb{r})$ into Eqs. \eqref{fn:temper} and \eqref{fn:phi}, the profiles of temperature and electric potential can be work out.
Therefore, the inhomogeneous plasma system with kappa stationary distribution can be solved completely by the above method including the electron density, the profiles of temperature and electric potential, and especially the kappa index which is usually a fitting parameter in other works.

\section{Simulations}
To verify the theory, we perform a full electrostatic PIC simulation in one dimension with periodic boundary conditions. 
In this numerical experiment, the standard momentum-conserving scheme of PIC simulation is used with a second-order interpolation in density and force weighting \cite{Birdsall2004}.
The electrostatic potential is solved from the Poisson equation \eqref{eq:P} by the method of fast Fourier transform. 
The phase space configurations of electrons are updated by the leap-frog algorithm \cite{Birdsall2004}, while the ions are set as a static background.

\begin{figure}[ht]
	\centering
    \subfigure[]
    {
    \includegraphics[width=0.47\textwidth]{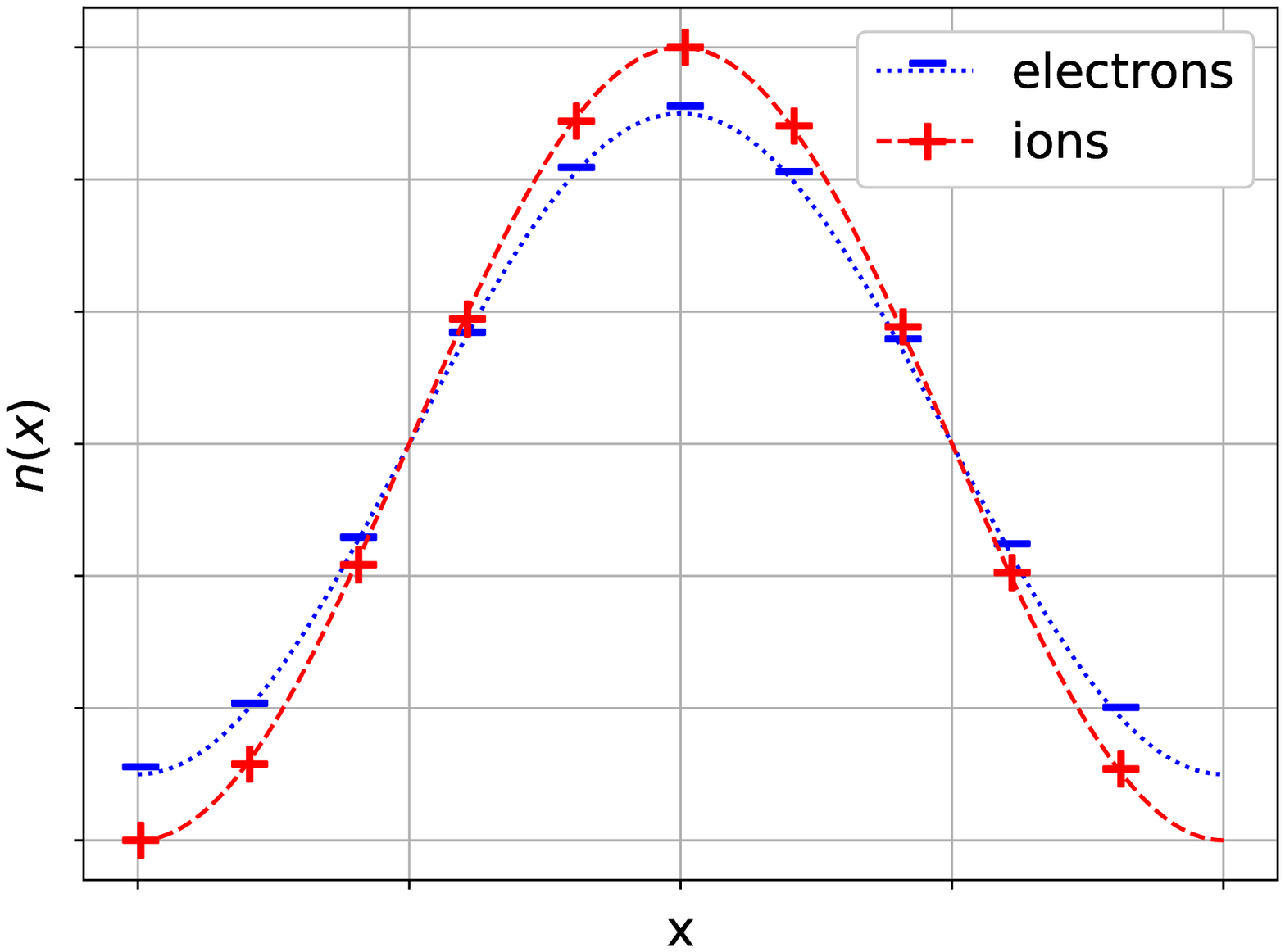}
    \label{fig:schematic-density}
    }
    \subfigure[]
    {
    \includegraphics[width=0.47\textwidth]{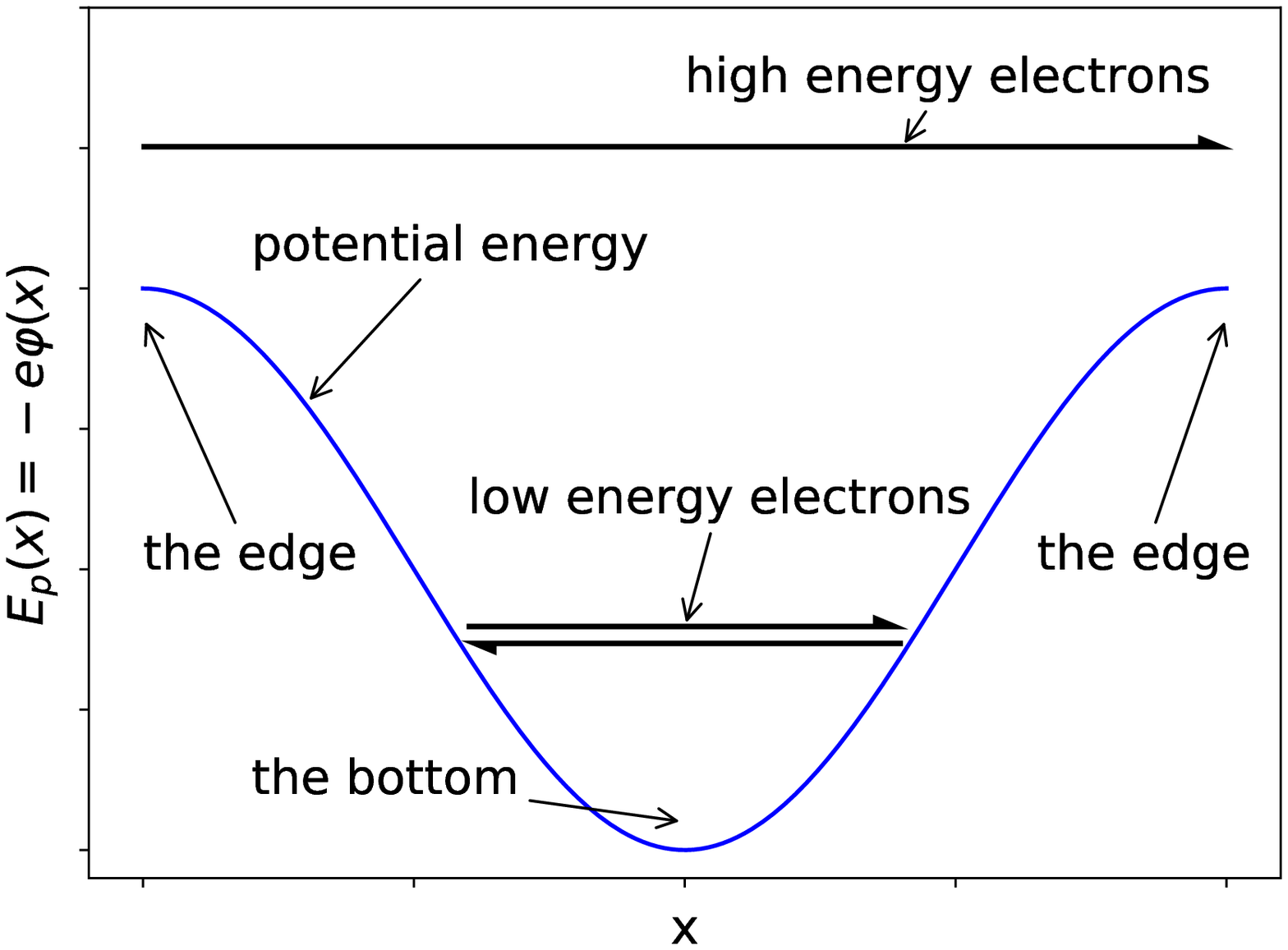}
    \label{fig:schematic-potential}
    }
    \caption{
        Schematic diagram of (a) density and (b) potential energy in the final stationary state.
    }
    \label{fig:schematic}
\end{figure}

To introducing the density inhomogeneity, we place the ions in a non-uniform manner, namely
\begin{equation}
    n_i(x) = n_0 \left(  1+\Delta_i \cos \frac{2 \pi x}{L}  \right),
    \label{fn:ni}
\end{equation}
where $\Delta_i$ is the maximum change in density and $L$ is the length of the simulated space.
At initial, the electrons are placed in the same manner of ions,
\begin{equation}
    n_e^{(init)}(x) = n_0 \left(  1+\Delta_e \cos \frac{2 \pi x}{L}  \right),
    \label{fn:ne}
\end{equation}
but with a larger maximum change of density, i.e. $\Delta_e > \Delta_i$.
The initial velocity distribution of electrons is set as isothermal Maxwellian.
It is worth noting that only the ion density is fixed as Eq. \eqref{fn:ni} in this simulation.
The profiles of the electron density and temperature, as well as the electron distribution, can evolve.
When the plasma system starts to evolve, the non-zero electric field and thermal pressure drive a standing wave with a finite amplitude.
In the evolution of the plasma, the wave is excited in the beginning, and then its amplitude is weakening due to Landau damping.
After a number of plasma periods, the wave is eased and the true solution of the Vlasov-Poisson system would evolve to a finer and finer structure in the velocity space.
However, in our PIC simulation, the velocity distribution will be described by a histogram with 20 bins.
It means that the distribution is only a coarse-grained distribution.
This coarse-grained velocity distribution of the collisionless plasma could evolve to a final stationary state, just like the Lynden-Bell's violent relaxation \cite{Levin2008a}. 
We will only focus on this coarse-grained velocity distribution in this study. 
At this final state, the variation range of the electron density should be slightly less than that of the ion density,
which is drawn in Fig. \ref{fig:schematic-density}.
The electric field caused by the local non-neutrality is balanced by the thermal pressure due to the inhomogeneous density and temperature.
At this moment, the low-energy electrons are trapped in the electric potential well but the high-energy electrons pass over the potential well as shown in Fig. \ref{fig:schematic-potential}.
The temperature at the bottom of the potential well is the averaged energy of both low- and high-energy electrons,
while at the edge of the potential well the temperature is only contributed from high-energy electrons.
Thereby, the temperature at the bottom of the potential well may be lower than that at the edge.
Considering that the electron density is large at the bottom of the potential well, one finds a negative correlation between density and temperature in the final steady state.
Therefore, the non-uniform density and temperature can be produced in our simulation.

In such a relaxation process, the initial standing wave has a finite amplitude and it would convert into a series of harmonics with different phase velocities, each of which flattens the velocity distribution function nearby its phase velocity.
Therefore, the electron velocity distribution is significantly changed by the nonlinear wave-particle interactions and then evolves towards the kappa distribution predicted by the theory.
Because the wave is completely eased, both of Landau damping and the nonlinear wave-particle interaction vanish in the final stationary state.

In the simulation, the arbitrary units are adopted for the parameters. 
We treat the Boltzmann constant $k_B$ and the vacuum dielectric constant $\varepsilon_0$ as one.
The plasma system evolves in the space of length $L$ which is divided into 200 cells with 5000 electrons per cell to reduce the numerical noises. 
For convenience, we let the plasma frequency $\omega_{pe} = \sqrt{n_0 e^2/(\varepsilon_0 k_B T_{init})}= 1$ as well as the initial temperature of electrons $T_{init}=1$, which induces that the magnitude of electron charge is $e = \sqrt{1/n_0}$.
Because of the long-range interactions between the particles, the collisionless relaxation becomes very slow, so the time of simulations are taken as $50000\omega_{pe}^{-1}$. 

\begin{figure}[ht]
	\centering
    \subfigure[]
    {
    \includegraphics[width=0.4\textwidth]{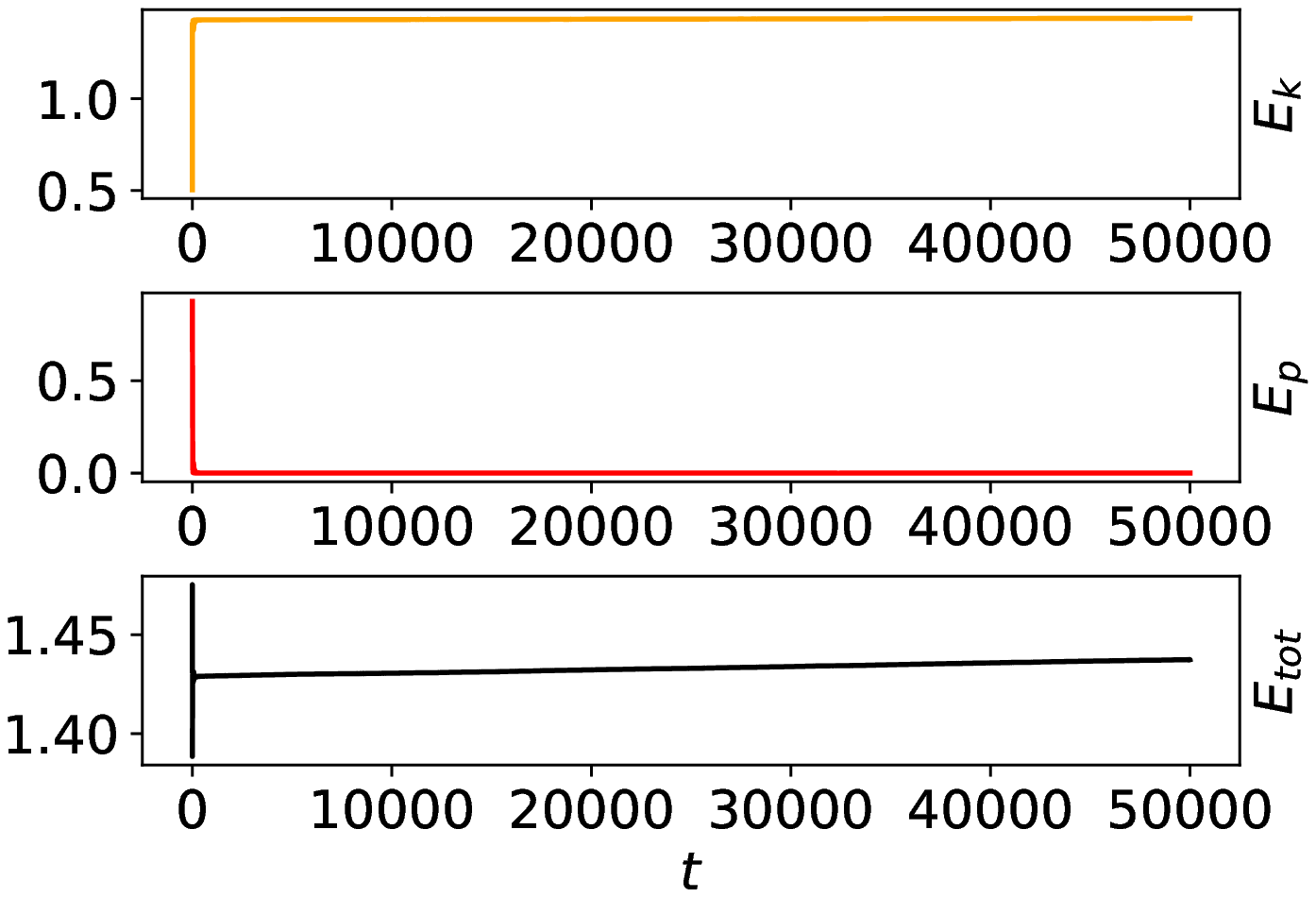}
    }
    \subfigure[]
    {
    \includegraphics[width=0.4\textwidth]{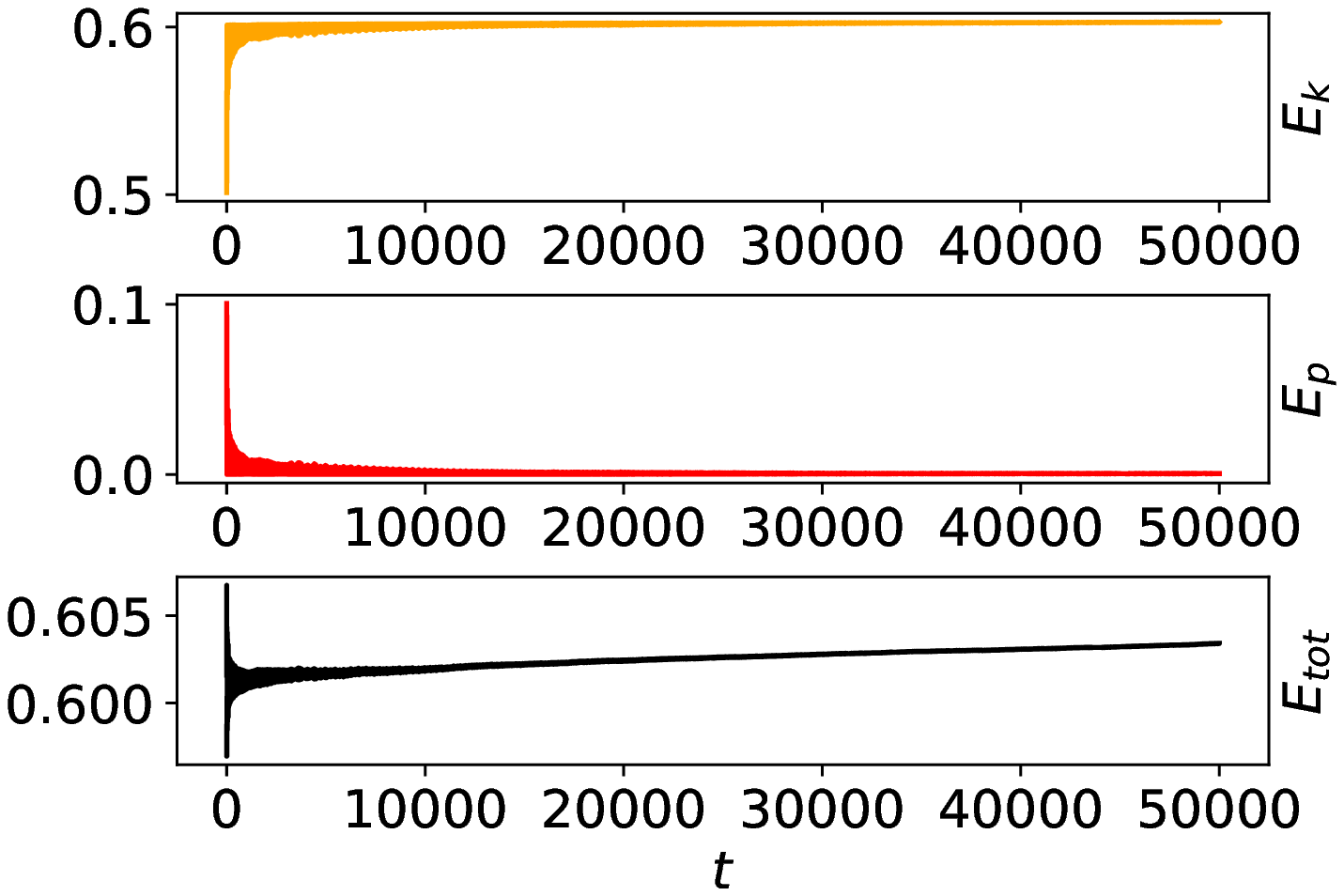}
    }
    \caption{
        The time evolution of energy, including the kinetic energy $E_k$, the potential energy $E_p$, and the total energy $E_{tot}$. 
        The simulations are made for two groups of parameters (a) $L = 60$, $\Delta_e=0.6$, $\Delta_i=0.4$ and (b) $L = 20$, $\Delta_e=0.4$, $\Delta_i=0.2$.
        It shows that in the comparison with the total runtime the wave is damped quickly.
    }
    \label{fig:energy}
\end{figure}

\begin{figure}[ht]
	\centering
    \subfigure[]
    {
    \includegraphics[width=0.4\textwidth]{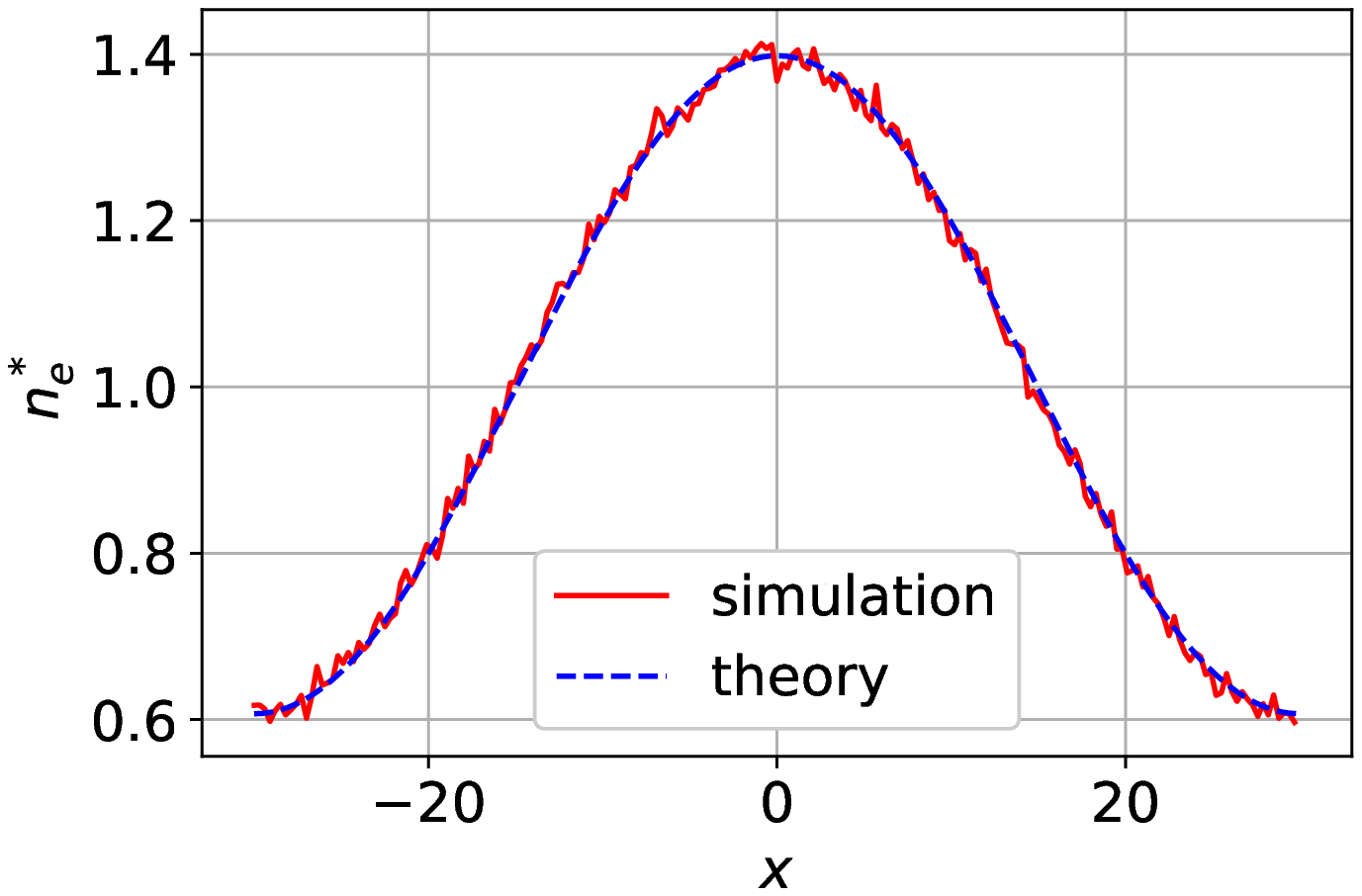}
    \label{fig:casea-nd}
    }
    \subfigure[]
    {
    \includegraphics[width=0.4\textwidth]{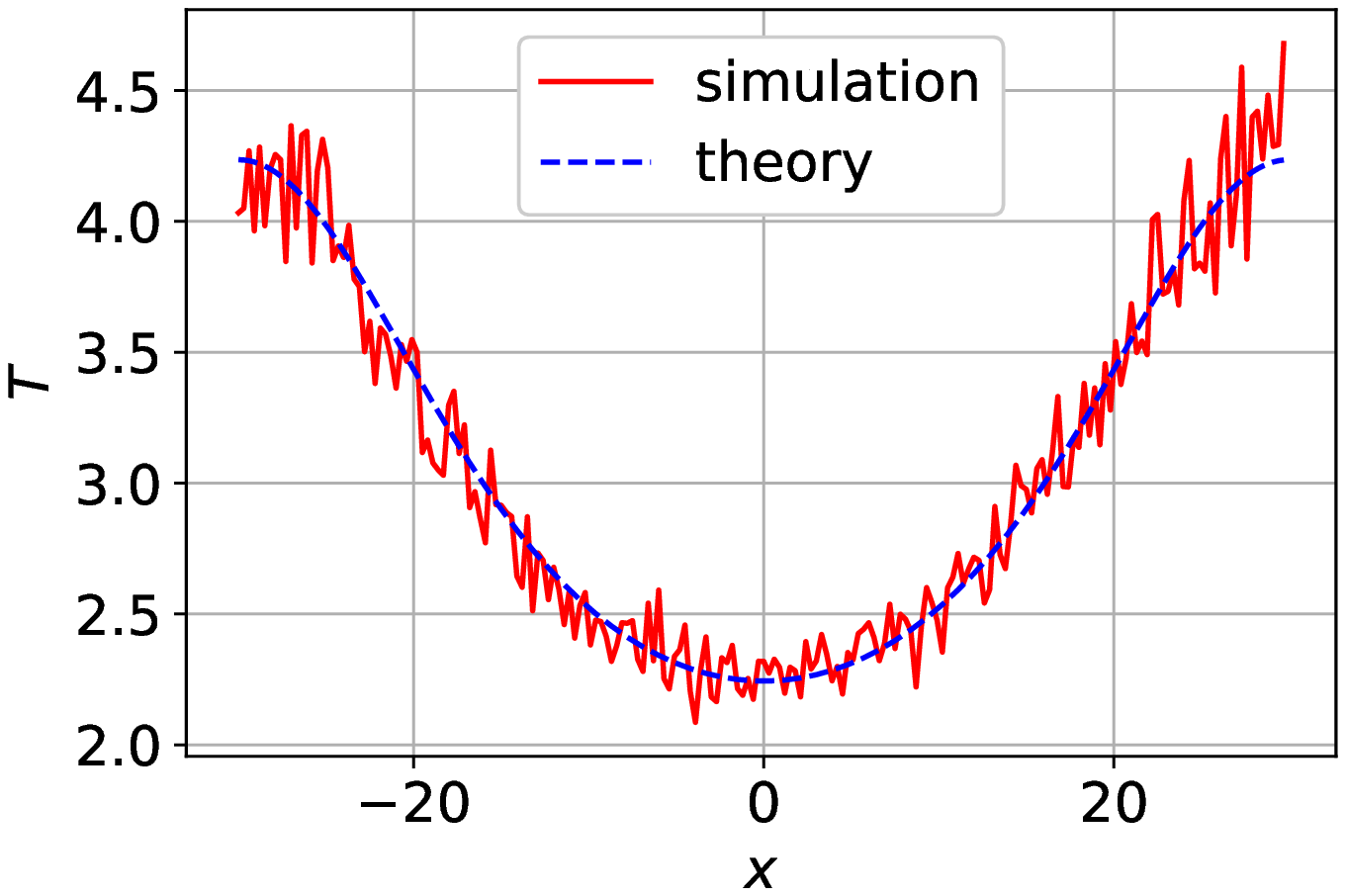}
    \label{fig:casea-td}
    }
    \subfigure[]
    {
    \includegraphics[width=0.4\textwidth]{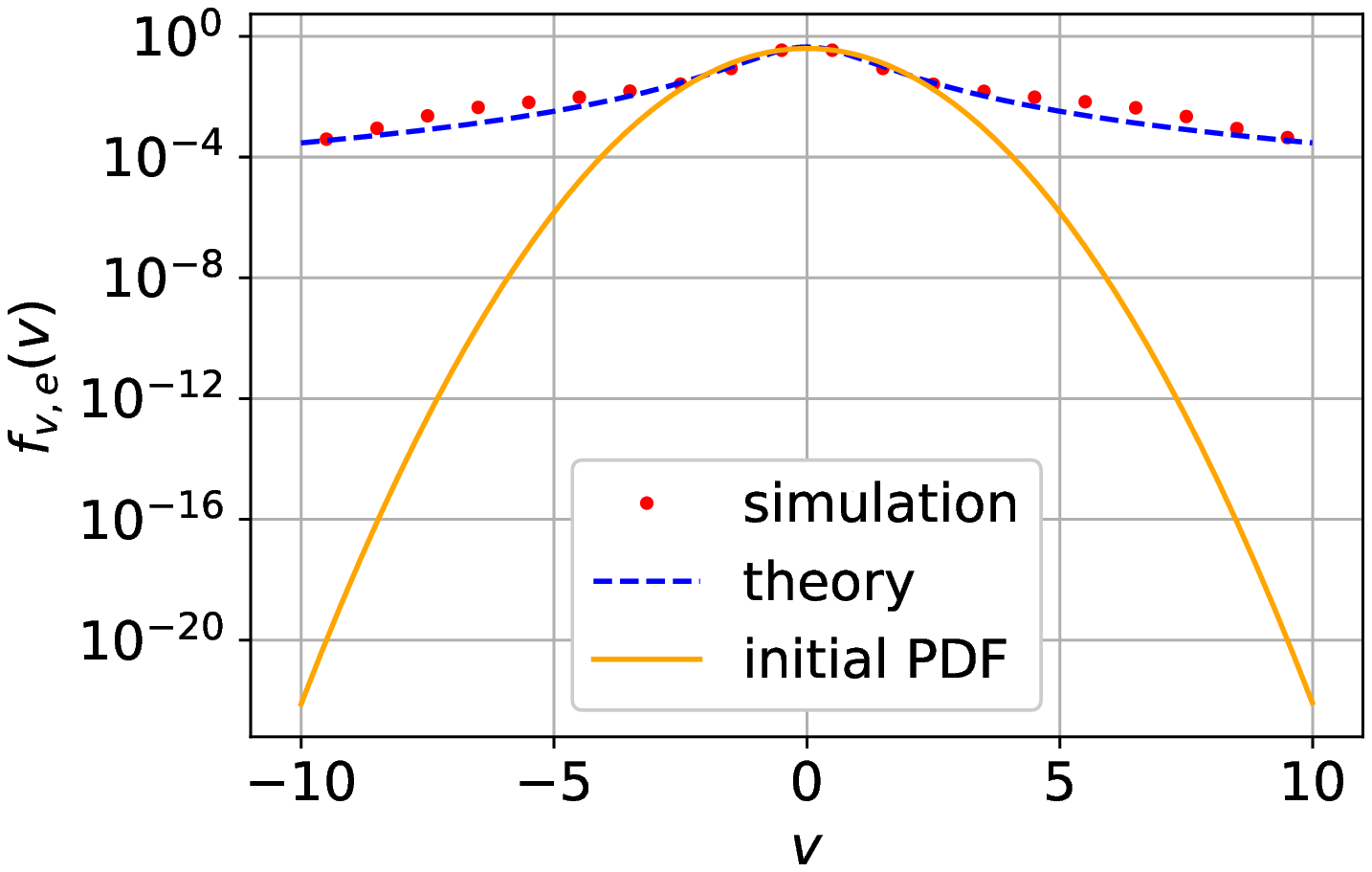}
    \label{fig:casea-vpdf}
    }
    \subfigure[] 
    {
    \includegraphics[width=0.4\textwidth]{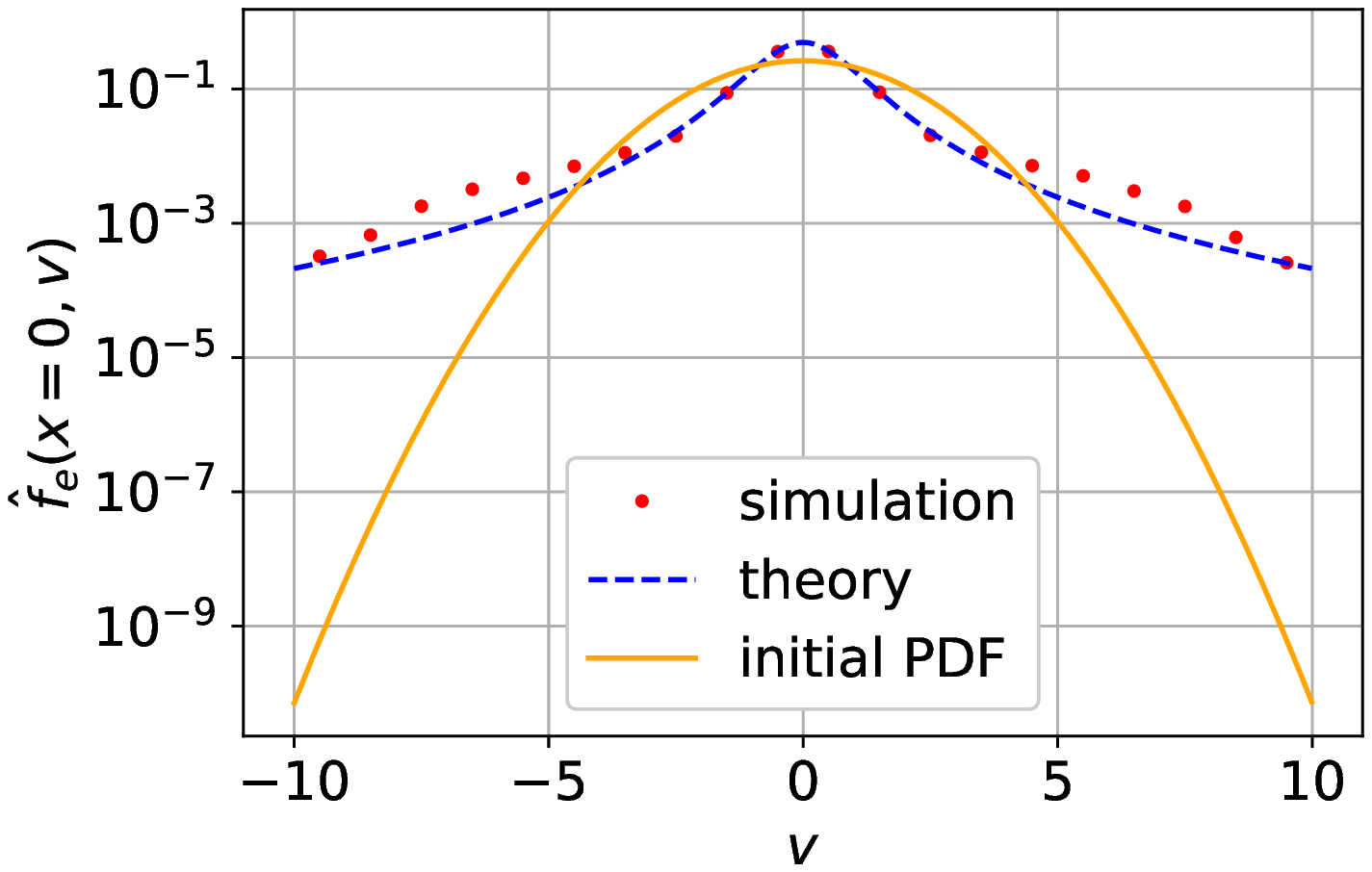}
    \label{fig:casea-lvpdf}
    }
	\caption{
        The simulation results of (a) the relative density, (b) the temperature distribution, (c) the global velocity distribution, and (d) the local velocity distribution at $x=0$ are drawn for the parameters $L = 60$, $\Delta_e=0.6$ and $\Delta_i=0.4$.
        It is solved from theoretical calculations that $\kappa_0=0.313970$ and $T_0= 2.89604$.
        In (a) and (b), the red solid line is the simulation result and the blue dashed line is the theoretical prediction.
        In (c) and (d), the red dotted line is the simulation result and the blue dashed line is the theoretical prediction.
        As a comparison, the initial velocity distribution is drawn as the orange solid line (i.e. the isothermal Maxwellian distribution).
}
    \label{fig:casea}
\end{figure}

\begin{figure}[ht]
	\centering
    \subfigure[]
    {
    \includegraphics[width=0.4\textwidth]{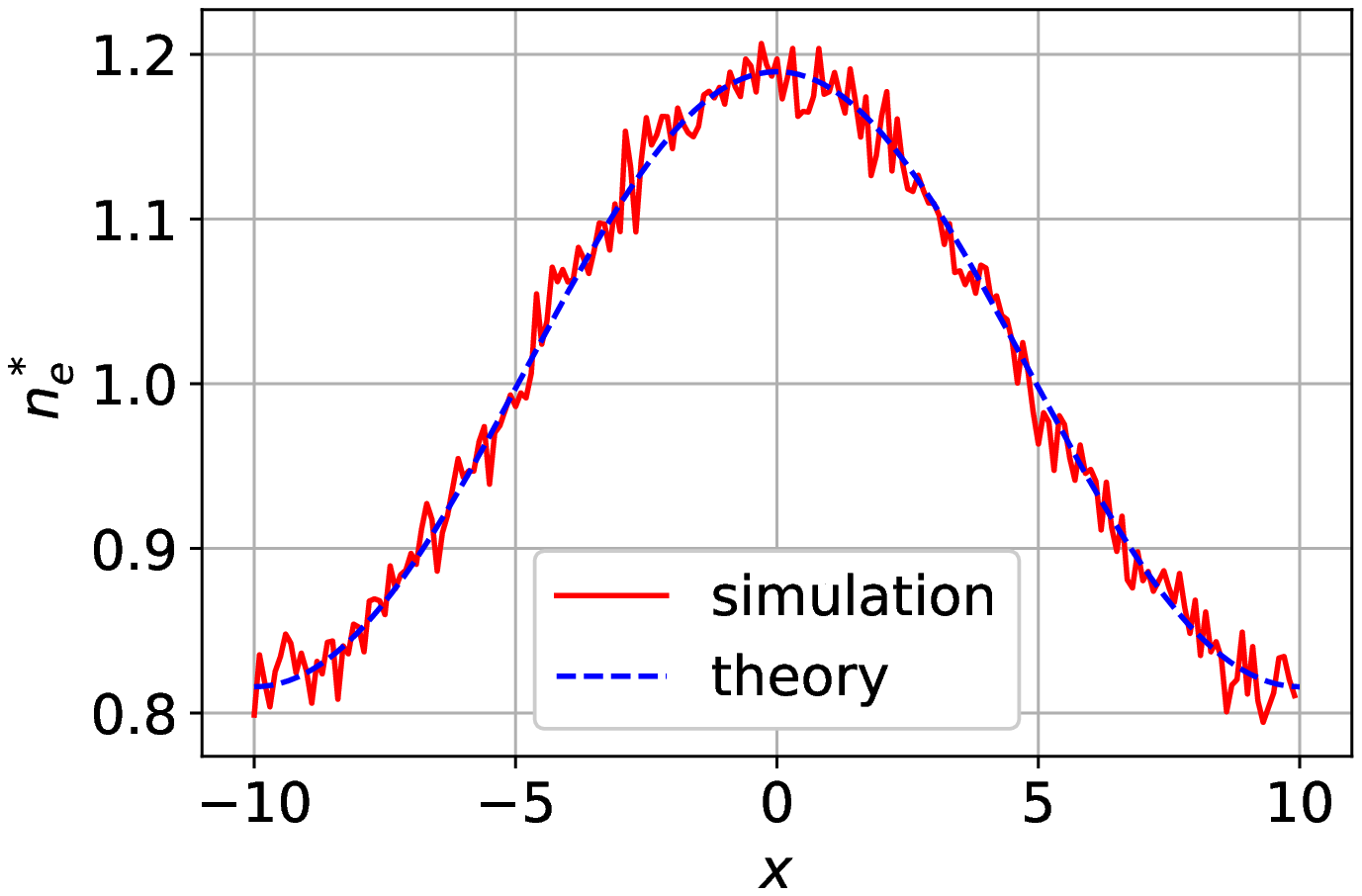}
    \label{fig:caseb-nd}
    }
    \subfigure[]
    {
    \includegraphics[width=0.4\textwidth]{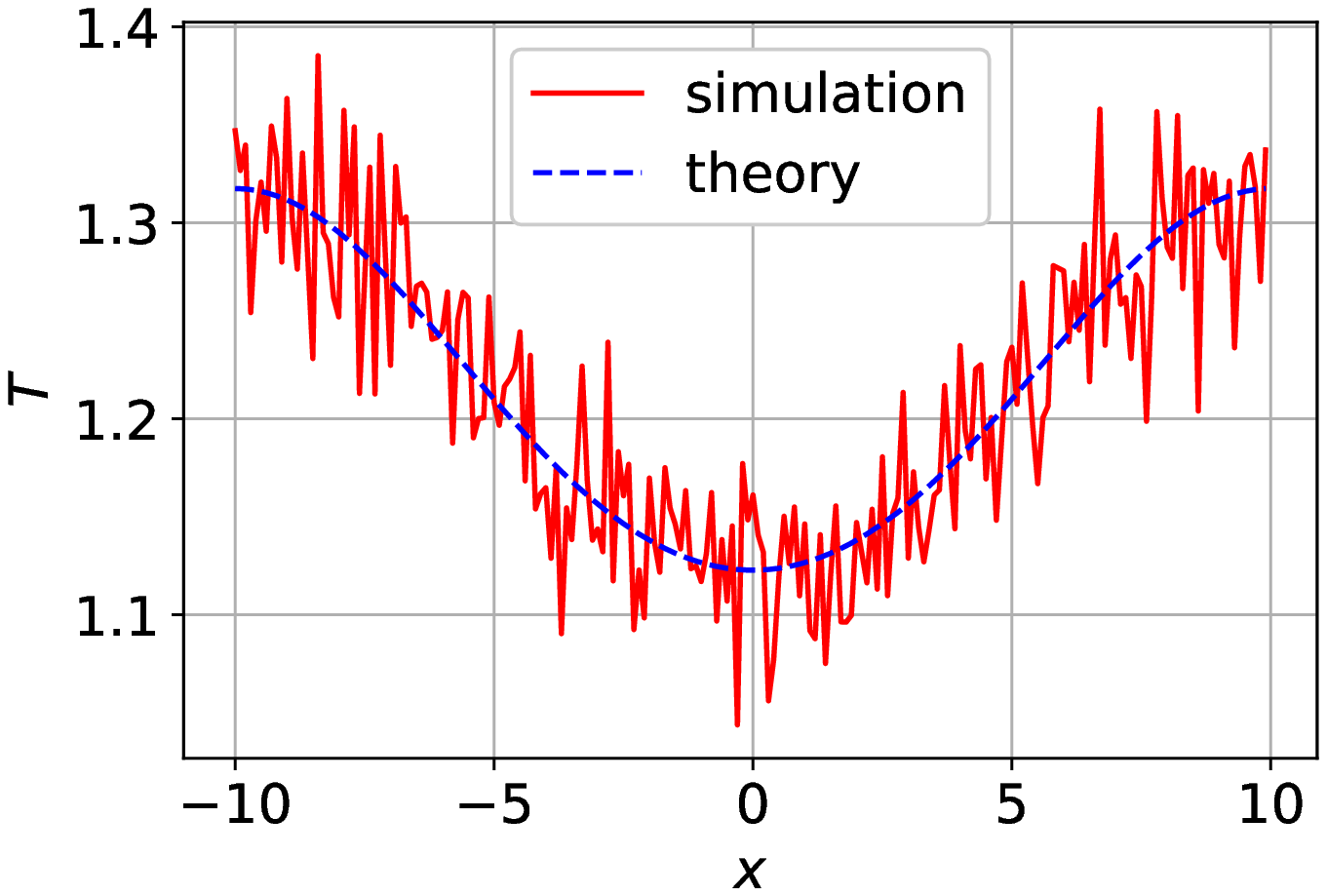}
    \label{fig:caseb-td}
    }
    \subfigure[]
    {
    \includegraphics[width=0.4\textwidth]{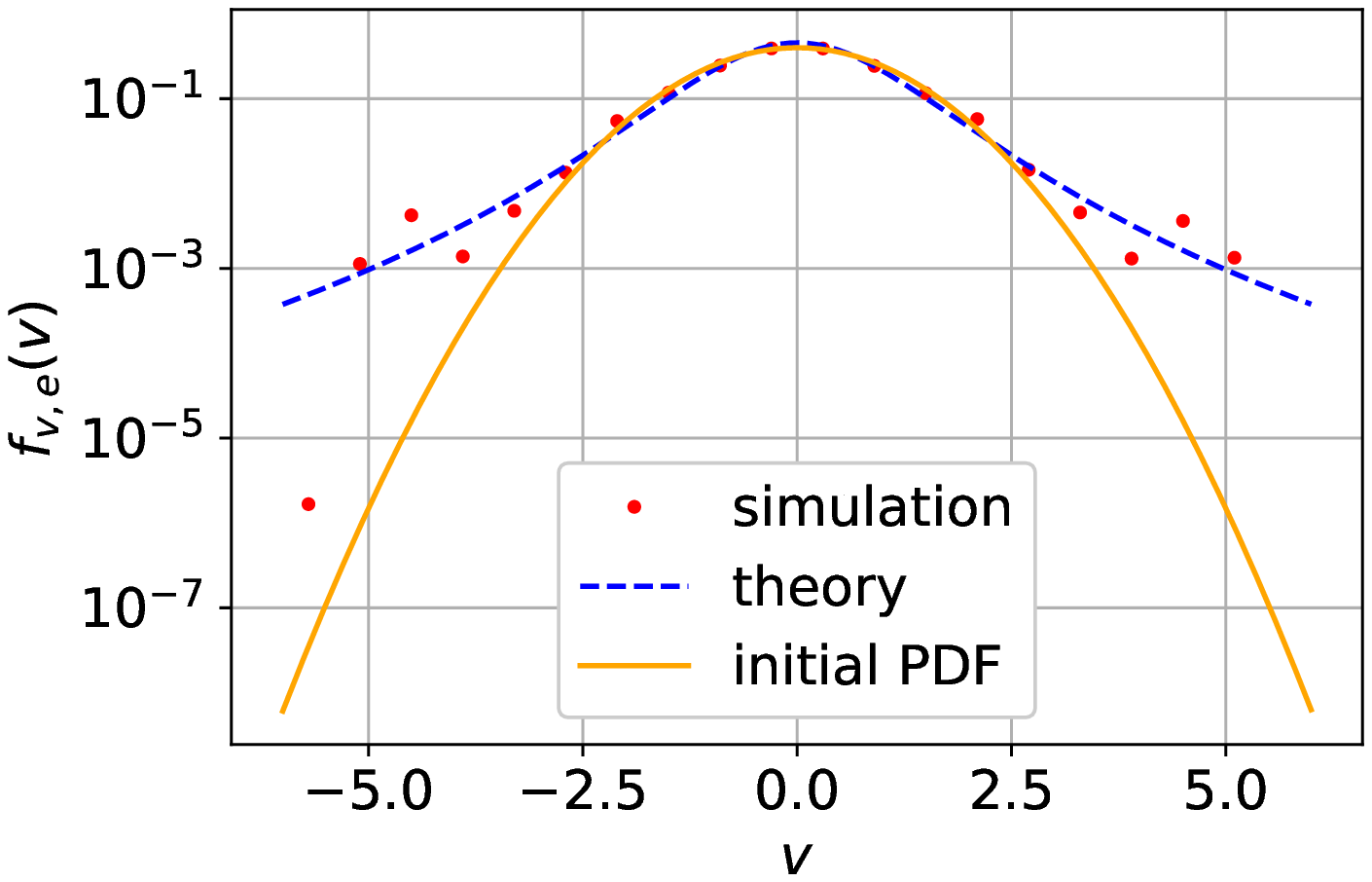}
    \label{fig:caseb-vpdf}
    }
    \subfigure[] 
    {
    \includegraphics[width=0.4\textwidth]{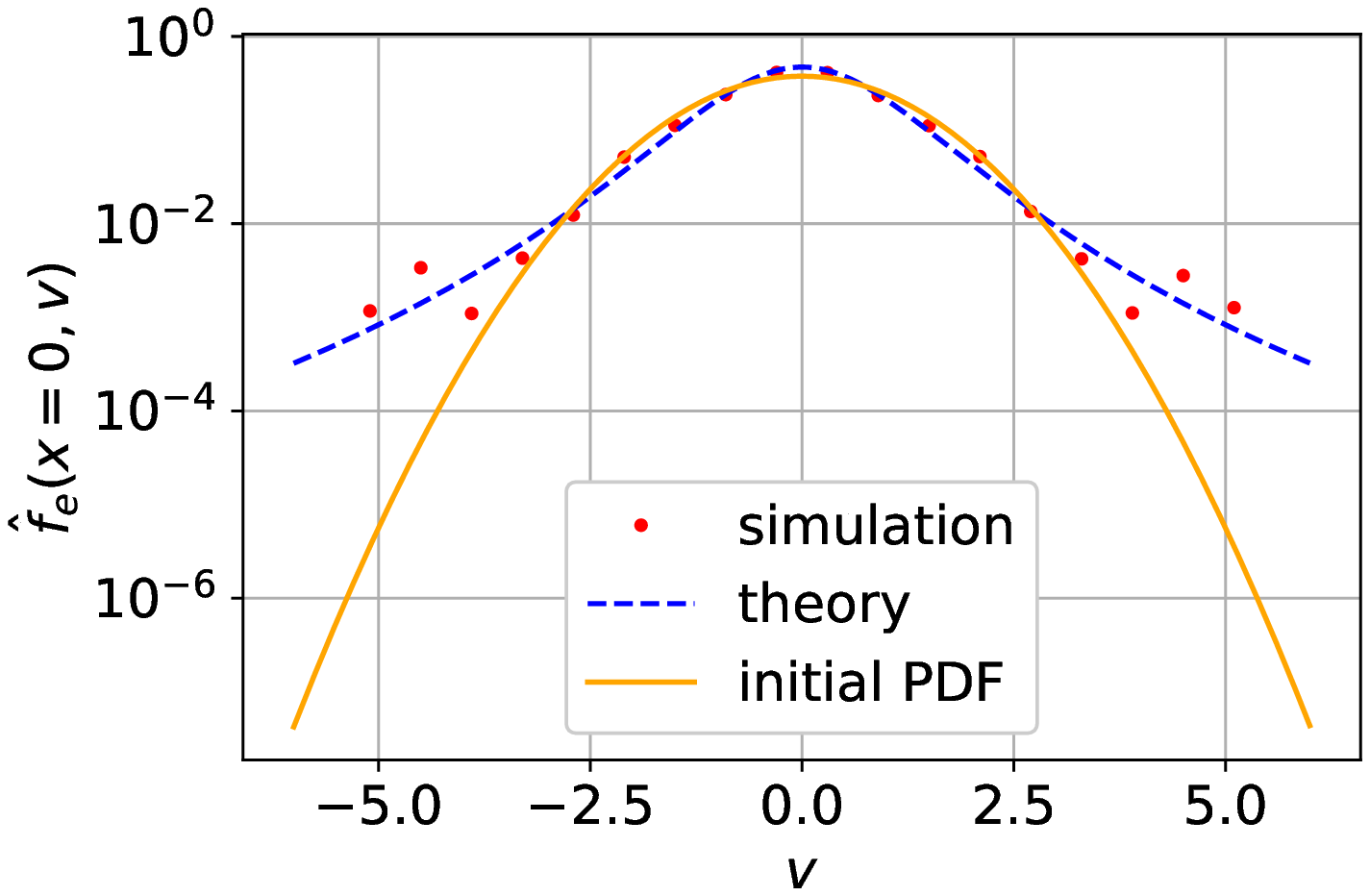}
    \label{fig:caseb-lvpdf}
    }
	\caption{
        The simulation results of (a) the relative density, (b) the temperature distribution, (c) the global velocity distribution, and (d) the local velocity distribution at $x=0$ are drawn for the parameters $L = 20$, $\Delta_e=0.4$ and $\Delta_i=0.2$.
        It is solved from theoretical calculations that $\kappa_0=1.35591$ and $T_0= 1.20854$.
        The legends in this figure are the same as those in Fig. \ref{fig:casea}.
}
    \label{fig:caseb}
\end{figure}
\begin{figure}[ht]
	\centering
    \subfigure[]
    {
    \includegraphics[width=0.4\textwidth]{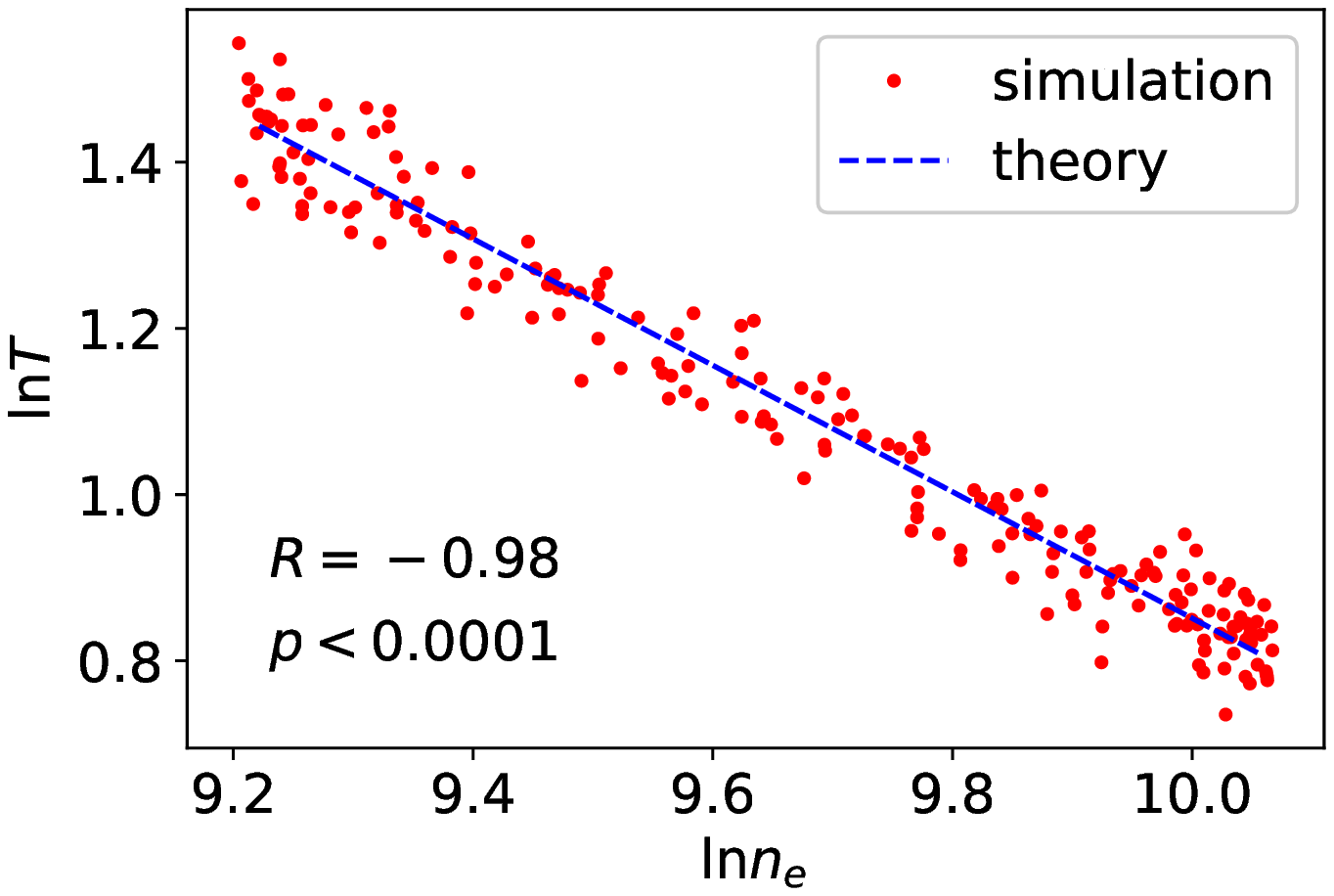}
    }
    \subfigure[]
    {
    \includegraphics[width=0.4\textwidth]{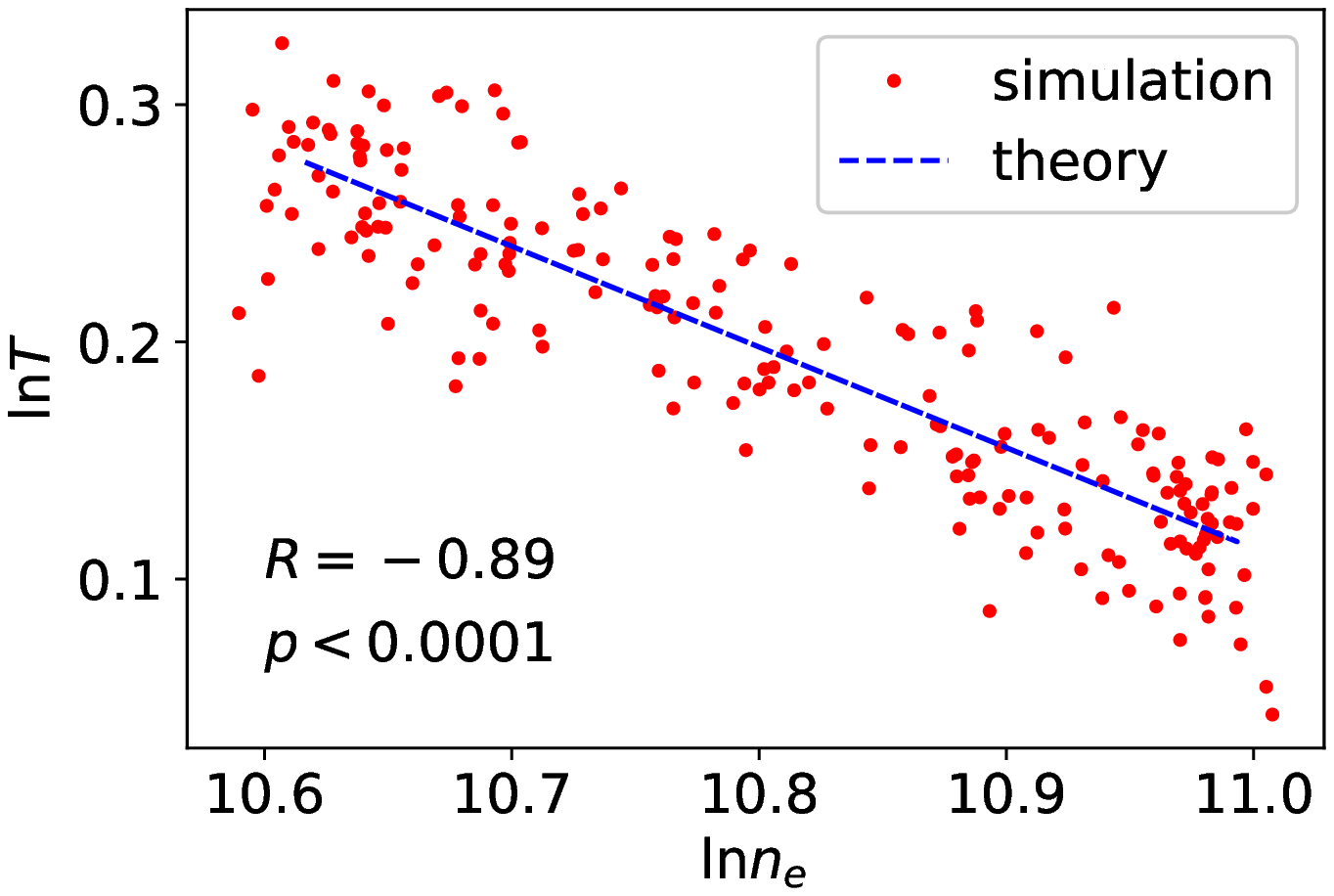}
    }
    \caption{
        The correlation between electron density $n_e$ and temperature $T$ with the parameters (a) $L = 60$, $\Delta_e=0.6$ and $\Delta_i=0.4$ and (b) $L = 20$, $\Delta_e=0.4$ and $\Delta_i=0.2$.
        The simulation result is drawn by the red dot, and the theoretical prediction is drawn by the blue dashed line.
        The linear correlation between $\ln n_e$ and $\ln T$ is studied by the Pearson correlation coefficient $R$ and the $p$ value which are shown in the figure. 
    }
    \label{fig:polytropic}
\end{figure}

The results of the simulations are drawn in Figs. \ref{fig:energy} - \ref{fig:polytropic}.
We perform two groups of simulations with different parameters, which are $L = 60$, $\Delta_e=0.6$, $\Delta_i=0.4$ denoted by Case A and $L = 20$, $\Delta_e=0.4$, $\Delta_i=0.2$ by Case B.
In the Figs. \ref{fig:casea} and \ref{fig:caseb}, the velocity distribution from the simulation is drawn as a histogram with 20 bins.
The theoretical predictions are calculated by the following steps.
Firstly, we need to determine the energy and entropy of the plasma system.
For a PIC simulation, both the energy and entropy increase in the time evolution of plasma due to numerical noises generated by unreal interactions between grids and macro-particles, which is usually called self-heating \cite{Birdsall2004}.
Therefore, the values of energy and entropy are calculated from the simulated final distribution, which results in more accurate values of $\kappa_0$ and $T_0$. 
The comparisons between the calculations from initial theoretical distribution and final simulated distribution are displayed in Appendix \ref{ap:com}.
Secondly, we numerically solve Eqs. \eqref{eq:u}, \eqref{fn:energy} and \eqref{fn:entropy} together with periodic boundary conditions
\begin{equation}
    n^*_e \left( -\frac{L}{2}  \right) = n^*_e \left( \frac{L}{2}  \right) ~~ \text{and} ~~ 
    \dv[]{n^*_e}{x}\bigg |_{x=-\frac{L}{2}} = \dv[]{n^*_e}{x}\bigg |_{x=\frac{L}{2}}
    \label{eq:u_bc}
\end{equation}
to calculate $n^*_e(x)$ and the critical parameters $\kappa_0$ and $T_0$.
Thirdly, by inserting these results, one gets the theoretical profiles of temperature \eqref{fn:temper} and potential \eqref{fn:phi}.
Substituting the temperature and density distribution back into \eqref{eq:pdf}, the electron distribution $f_e(\vb{r},\vb{v})$ can be obtained.
After that, the global velocity distribution of electrons $f_{v,e}(v)$ can be calculated from the integral,
\begin{equation}
    f_{v,e}(v) = \frac{1}{N} \int_{- \frac{L}{2} } ^{ \frac{L}{2}  } f_e( x, v ) \dd{ x }.
\end{equation}

The errors between simulations and theoretical results can be measured by
\begin{equation}
    \chi_P^2 = \int \frac{ \left[ P^{sim}(\epsilon)-P^{theo}(\epsilon) \right] ^2 }{P^{theo}(\epsilon)} \dd{\epsilon},
\end{equation}
where $P(\epsilon)$ is the physical quantity to be examined. 
We estimate the errors for relative number density $\chi_n^2$, the temperature $\chi_T^2$, the local velocity distribution $\chi_{fl}^2$, and the global velocity distribution $\chi_{fg}^2$, which are shown in Table. \ref{tab:err}.

\begin{table}[h]
\centering
\begin{tabular}{lllll}
    \hline
    \hline
           & $\chi_n^2$~~~ & $\chi_T^2$~~~ & $\chi_{fl}^2$~~~ & $\chi_{fg}^2$~~~ \\
   \hline
    Case A~~~ & $1.07 \times 10^{-2}$~~~ & $3.65 \times 10^{-1}$~~~ & $3.73 \times 10^{-2}$~~~ & $5.30 \times 10^{-2}$~~~\\
    Case B~~~ & $4.05 \times 10^{-3}$~~~ & $2.04 \times 10^{-2}$~~~ & $1.74 \times 10^{-2}$~~~ & $2.09 \times 10^{-2}$~~~\\
    \hline
    \hline
\end{tabular}
    \caption{The errors between simulations and theoretical predictions.}
    \label{tab:err}
\end{table}

The simulation results show excellent agreement with the theoretical predictions for the profiles of relative density $n^*_e(x)$ in Figs. \ref{fig:casea-nd} and \ref{fig:caseb-nd}, as well as the temperature distribution $T(x)$ in Figs. \ref{fig:casea-td} and \ref{fig:caseb-td}.
For the velocity distribution in Figs. \ref{fig:casea-vpdf}, \ref{fig:casea-lvpdf}, \ref{fig:caseb-vpdf} and \ref{fig:caseb-lvpdf}, the simulated solutions fluctuate around the theoretical results to a small extent.   
Nevertheless, compared to the initial Maxwellian distribution, the final velocity distribution is much closer to the predicted kappa distribution with no doubts.
We must explain that the simulation results are not caused by numerical dissipation because these results are consistent with the theoretical predictions not only for the velocity distribution but also the profiles of density and temperature with different parameters.
If the velocity distribution is caused by numerical diffusion, the non-uniform temperature profile may not appear.
Even if the non-uniform temperature appears in the case of numerical diffusion, the profile of temperature should not be consistent with the theoretical prediction.

Besides, we analyze the linear correlation between $\ln n_e$ and $\ln T$ in Fig. \ref{fig:polytropic}.
The results of the Pearson correlation coefficient $R$ and the $p$ value indicate a high linear correlation between $\ln n_e$ and $\ln T$.
That is to say, the polytropic EOS is a very good model to describe the relationship between the density and temperature in our simulation.

\section{Influence of different initial states}
In the previous sections, we find that the kappa index of the final stationary distribution relies on the initial state of plasma.
To analyze the influence of different initial states, we can calculate the kappa index of the final stationary state by numerically solving Eqs. \eqref{eq:u}, \eqref{fn:energy} and \eqref{fn:entropy}, instead of doing PIC simulations.
However, the plasma system is so complex that we cannot conduct the research over all different initial states.
Therefore, we select three dominant parameters to distinguish initial states, namely the space length $L$, the amplitude of initial density of electrons $\Delta_e$ and ions $\Delta_i$ in Eqs. \eqref{fn:ne} and \eqref{fn:ni}.
In order to study the influence of these three parameters, we calculate the kappa index in the following conditions:
\begin{enumerate}
    \item Only varying the space length $L$.
    \item Only varying the amplitude of electrons $\Delta_e$.
    \item Fixing the difference $\Delta_e-\Delta_i$ and varying $\Delta_e$ and $\Delta_i$ concurrently.
\end{enumerate}

The results are presented in Table \ref{tab:var-l} - \ref{tab:var-dedi}.
Generally speaking, if the profile of ion density is the same, a larger initial potential energy leads to a larger temperature inhomogeneity in the final stationary state, 
and a smaller kappa index is expected due to the polytropic EOS.
On the other hand, with the same potential energy at initial, the kappa index should be small with a small initial entropy because of the properties of the kappa distribution function.
\begin{table}[h]
\centering
\begin{tabular}{llll}
    \hline
    \hline
    $L$     & $E$          & $S$          & $\kappa_0$ \\
    \hline
    $5$     & $0.506333$   & $2.933737$   & $1.014094$  \\
    $10$    & $0.525330$   & $3.626884$   & $0.984514$ \\
    $20$    & $0.601321$   & $4.320031$   & $0.657696$ \\
    $40$    & $0.905285$   & $5.013179$   & $0.282033$ \\
    $60$~~  & $1.411891$~~ & $5.418644$~~ & $0.147081$ \\
    \hline
    \hline
\end{tabular}
    \caption{Calculations of the kappa index with variable space length $L$, fixed $\Delta_e=0.6$, and fixed $\Delta_i=0.4$.}
    \label{tab:var-l}
\end{table}

\begin{table}[h]
\centering
\begin{tabular}{llll}
    \hline
    \hline
    $\Delta_e$& $E$         & $S$          & $\kappa_0$ \\
    \hline
    $0.5$    & $0.525330$   & $4.350033$   & $1.451663$  \\
    $0.6$    & $0.601321$   & $4.320031$   & $0.657696$ \\
    $0.7$    & $0.727973$   & $4.283048$   & $0.360448$ \\
    $0.8$    & $0.905285$   & $4.237814$   & $0.215948$ \\
    $0.9$~~  & $1.133257$~~ & $4.181923$~~ & $0.135597$ \\
    \hline
    \hline
\end{tabular}
    \caption{Calculations of the kappa index with variable $\Delta_e$, fixed $\Delta_i=0.4$, and fixed length $L=10$.}
    \label{tab:var-de}
\end{table}

\begin{table}[h]
\centering
\begin{tabular}{llll}
    \hline
    \hline
    $\Delta_e$& $E$         & $S$          & $\kappa_0$ \\
    \hline
    $0.1$    & $0.601321$   & $4.391910$   & $0.799256$ \\
    $0.2$    & $0.601321$   & $4.373825$   & $0.746759$ \\
    $0.3$    & $0.601321$   & $4.350033$   & $0.700250$ \\
    $0.4$    & $0.601321$   & $4.320031$   & $0.657696$ \\
    $0.5$~~  & $0.601321$~~ & $4.283048$~~ & $0.617006$ \\
    \hline
    \hline
\end{tabular}
    \caption{Calculations of the kappa index with variable $\Delta_e$, fixed $\Delta_e-\Delta_i=0.2$, and fixed length $L=10$.}
    \label{tab:var-dedi}
\end{table}

For these reasons, the results in Table \ref{tab:var-l} - \ref{tab:var-dedi} can be explained well.  
Firstly, in Table \ref{tab:var-l}, we find that during the increment of length $L$, the kappa index decreases, while both energy and entropy increase.
Although the large energy and entropy take the opposite effects on the value of kappa in terms of the above analysis, the calculation results indicate that the energy has the dominant influence on the kappa index.
Secondly, it is shown in Table \ref{tab:var-de} that a larger $\Delta_e$, namely a larger initial potential energy, and a smaller initial entropy would have the same effects and result in a smaller kappa index for the same reason.
Thirdly, Table \ref{tab:var-dedi} implies that the kappa index decreases as the initial entropy decreases. This is also in accordance with our analysis.

\section{Summary}
In conclusion, we study the stationary state of an inhomogeneous isotropic plasma.
In the Vlasov-Poisson system, we find that the stationary electron distribution follows the kappa distribution \eqref{eq:pdf} under the polytropic EOS \eqref{eq:poly-rela} and the assumption on the form of the local velocity distribution \eqref{assump:argu} . 
In the result \eqref{eq:pdf}, the local velocity distribution is analytically derived, while the electron density could be numerically solved from Eq. \eqref{eq:u} with certain boundary conditions if the ion density is given.
The profiles of temperature \eqref{fn:temper} and electrostatic potential \eqref{fn:phi} can both be calculated according to the numerical solution of Eq. \eqref{eq:u}.
If the energy and the entropy are given by a certain initial state, the parameters $\kappa_0$ and $T_0$ can be determined by solving Eqs. \eqref{eq:u}, \eqref{fn:energy} and \eqref{fn:entropy} at the same time.
Thus, in our theory, there are no fitting parameters.
In addition, we make the PIC simulation to examine the theory, the results of which show excellent agreement with the theoretical predictions.
At last, the influence of different initial states is discussed.
It suggests that a larger initial potential energy and a smaller initial entropy would lead to a smaller kappa index in the final steady state.

The kappa distribution in this paper is derived from the collisionless Vlasov-Poisson equations with the polytropic EOS.
It is one of the mechanisms generating the kappa distribution.
It is worth to note that there is no conflict between our work and Refs. \cite{Hasegawa1985PRL,Ma1998GRL,Vocks2003AJ,Bian2014AJ,Shizgal2018PRE} in which the kappa distribution is derived from the collisional Fokker-Planck equation. 
The reason is that the plasma studied in Refs. \cite{Hasegawa1985PRL,Ma1998GRL,Vocks2003AJ,Bian2014AJ,Shizgal2018PRE} is different from that in our work.
The studies of Refs. \cite{Hasegawa1985PRL,Ma1998GRL,Vocks2003AJ,Bian2014AJ,Shizgal2018PRE} focus on the collisional isothermal plasma while we study the collisionless plasma with non-uniform density and temperature.
The mechanism generating the kappa distribution can be different in different plasmas.
Therefore, all of them are possible explanations of the formation of kappa distribution.
On the other hand, the in-depth correlation between different mechanisms generating kappa distribution is a very interesting but difficult question which is still unknown.
We will study this topic in the future. 

As a final remark, the polytropic EOS is usually used as an approximation to describe the relationship between density and temperature in space plasmas.
In our model, the theoretical results are derived rigorously from the polytropic assumption. 
Therefore, the accuracy of the theoretical results depends on the accuracy of the polytropic EOS.
The results in Fig. \ref{fig:polytropic} suggest that the polytropic EOS is at least an excellent approximation in our simulation.
Some observations \cite{Sittler1980,Belmont1992,MeyerVernet1995,Pang2016} also support that the polytropic EOS is highly credible in some plasmas.

\begin{acknowledgments}
This work was supported by the National Natural Science Foundation of China (No.11775156) and by the Fundamental Research Funds for the Central Universities, Civil Aviation University of China (No.3122019138).
The author is also grateful for many thoughtful and valuable comments from the anonymous referees.
\end{acknowledgments}

\section*{Data Availability}
The data that support the findings of this study are available from the corresponding author upon reasonable request.

\appendix
\section{The proof of the identity $T \dd{\ln A} / \dd{T} = -d/2$}
\label{ap:equality}
The normalization of the local velocity distribution function can be expressed as,
\begin{equation}
    A[T(\vb{r})] = \left\{\int g \left[\frac{mv^2}{2k_B T(\vb{r})}\right] \dd{\vb{v}} \right\}^{-1}.
\end{equation}
The $d$-dimensional differential volume element $\dd{\vb{v}}$ can be rewritten as a product of the surface area of a $(d-1)$-dimensional sphere with $\dd{v}$, namely,
\begin{equation}
\dd{\vb{v}} = \frac{2\pi^{\frac{d}{2}}}{\Gamma\left(\frac{d}{2}\right)}v^{d-1} \dd{v},
\label{dev:dv}
\end{equation}
which leads to the normalization, 
\begin{equation}
    A[T(\vb{r})] = \left\{\int_{0}^{\infty} g \left[ \frac{mv^2}{2k_B T(\vb{r})}  \right] \frac{2 \pi^{ \frac{d}{2} }}{\Gamma \left( \frac{d}{2} \right)} v^{d-1}  \dd{v} \right\}^{-1}.
    \label{dev:norm1}
\end{equation}
Let $W^* = \frac{mv^2}{2k_BT(\vb{r})}$, then Eq. \eqref{dev:norm1} turns out to be,
\begin{equation}
    A[T(\vb{r})] = \Gamma \left( \frac{d}{2}  \right) \left[ \frac{2 \pi k_B T(\vb{r})}{m}  \right]^{- \frac{d}{2} } I_1^{-1},
\end{equation}
with the integral $I_1$,
\begin{equation}
    I_1 = \int_0^{\infty} g(W^*) (W^*)^{ \frac{d}{2} -1 } \dd{W^*}.
\end{equation}
Noting that $I_1$ is an integral constant independent of $T$, we can calculate, 
\begin{equation}
    \dv[]{\ln A}{T} = \frac{1}{A} \dv[]{A}{T} = - \frac{d}{2 T},
\end{equation}
and therefore
\begin{equation}
    T\dv[]{\ln A}{T} = - \frac{d}{2}.
\end{equation}

\section{The consistency between the kappa solution and the local temperature}
\label{ap:consistency}
Substituting the solution \eqref{fun:kappaPDF} into the right side of Eq. \eqref{def:T}, the integral can be written as,
\begin{align}
I_2 = & \frac{1}{2} m \int v^2 \hat{f}_e(\vb{r},\vb{v}) \dd{\vb{v}} \notag \\
  = & \frac{1}{2} m \left[ \frac{m}{2 \pi k_B T(\vb{r}) \kappa_0} \right]^ \frac{d}{2}  \frac{\Gamma(\kappa_0+1+ \frac{d}{2})}{\Gamma(\kappa_0+1)} \int v^2 \left[ 1 + \frac{1}{\kappa_0} \frac{mv^2}{2k_BT(\vb{r})} \right]^{-\kappa_0-1-\frac{d}{2}} \dd{\vb{v}}.
\end{align}
By using Eq. \eqref{dev:dv}, the integral $I_2$ becomes,
\begin{equation}
I_2 = \frac{1}{2} m \left[ \frac{m}{2 \pi k_B T(\vb{r}) \kappa_0} \right]^ \frac{d}{2}  \frac{\Gamma(\kappa_0+1+ \frac{d}{2})}{\Gamma(\kappa_0+1)} \frac{2\pi^{\frac{d}{2}}}{\Gamma\left(\frac{d}{2}\right)} \int_0^\infty v^{d+1} \left[ 1 + \frac{1}{\kappa_0} \frac{mv^2}{2k_BT(\vb{r})} \right]^{-\kappa_0-1-\frac{d}{2}} \dd{v}.
\end{equation}
If we denote $t= \frac{1}{\kappa_0} \frac{mv^2}{2k_BT(\vb{r})}$, one has
\begin{equation}
    \dd{v} = \frac{1}{2} \left[\frac{\kappa_0 2 k_B T(\vb{r})}{m} \right]^{\frac{1}{2}} t^{-\frac{1}{2}} \dd{t}.
    \label{dev:dt}
\end{equation}
With Eq. \eqref{dev:dt}, the integral $I_2$ can be calculated,
\begin{align}
  I_2 = & \frac{1}{2} m \left[ \frac{m}{2 \pi k_B T(\vb{r}) \kappa_0} \right]^ \frac{d}{2}  \frac{\Gamma(\kappa_0+1+ \frac{d}{2})}{\Gamma(\kappa_0+1)}  \frac{\pi^{\frac{d}{2}}}{\Gamma\left(\frac{d}{2}\right)} \int_0^\infty \left[\frac{\kappa_0 2 k_B T(\vb{r})}{m} \right]^{\frac{d}{2}+1} t^{\frac{d}{2}} (1 + t)^{-\kappa_0-1-\frac{d}{2}} \dd{t} \notag \\
      = & \kappa_0  k_B T(\vb{r}) \frac{\Gamma(\kappa_0+1+ \frac{d}{2})}{\Gamma(\kappa_0+1)} \frac{1}{\Gamma\left(\frac{d}{2}\right)} \int_0^\infty  t^{\frac{d}{2}} (1 + t)^{-\kappa_0-1-\frac{d}{2}} \dd{t} \notag \\
      = & \kappa_0  k_B T(\vb{r}) \frac{\Gamma(\kappa_0+1+ \frac{d}{2})}{\Gamma(\kappa_0+1)} \frac{1}{\Gamma\left(\frac{d}{2}\right)} \frac{ \Gamma(\frac{d}{2}+1) \Gamma(\kappa_0)} {\Gamma(\frac{d}{2}+1+\kappa_0)} \notag \\
      = & \frac{d}{2} k_BT(\vb{r}),
\end{align}
where the formula of the beta integral\cite{Olver2010} has been used.

\section{The verification of the solution of the Vlasov-Poisson system}
In this section, we will prove that if $f_e$ is given by Eq. \eqref{eq:pdf} with $n_e$ taken from Eq. \eqref{eq:u}, then $f_e$ is a stationary solution of the Vlasov-Poisson system \eqref{eq:V} and \eqref{eq:P}.
We start from Eq. \eqref{eq:u}. This equation can be rearranged as,
\begin{equation}
    \frac{\kappa_0}{\kappa_0+1} \frac{k_B T_0}{e} \nabla \cdot \left( {n_e^*}^{-\frac{\kappa_0+2}{\kappa_0+1}} \nabla n_e^* \right) =  \frac{n_0 e}{\epsilon_0} (n_e^*-n_i^*).
    \label{ap:P-dev}
\end{equation}
By comparing Eq. \eqref{ap:P-dev} with the Poisson equation \eqref{eq:P}, one can find, 
\begin{equation}
    \nabla \varphi = \frac{\kappa_0}{\kappa_0+1} \frac{k_B T_0}{e} {n_e^*}^{-\frac{\kappa_0+2}{\kappa_0+1}} \nabla n_e^*.
    \label{ap:dphi}
\end{equation}
The partial derivatives of $f_e$ can be calculated directly from Eq. \eqref{eq:pdf},
\begin{align}
\nabla f_e=& \frac{\nabla n_e^*}{n_e^*}f_e 
    + \frac{d}{2} \frac{1}{\kappa_0+1} \frac{\nabla n_e^*}{n_e^*}f_e 
    - (\kappa_0+1+\frac{d}{2}) 
        \frac{ 
            \frac{1}{\kappa_0+1} \frac{1}{\kappa_0} \frac{mv^2}{2k_B T_0}{n_e^*}^{\frac{1}{\kappa_0+1}} 
            }{
            1+\frac{1}{\kappa_0} \frac{mv^2}{2k_B T_0}{n_e^*}^{\frac{1}{\kappa_0+1}}
            }
            \frac{\nabla n_e^*}{n_e^*}f_e \notag \\
          =& \frac{\nabla n_e^*}{n_e^*}f_e 
          \left[
    1 + \frac{d}{2} \frac{1}{\kappa_0+1} -\left(1+\frac{d}{2}\frac{1}{\kappa_0+1}\right) 
    \left(
        1-\frac{1}{
            1+\frac{1}{\kappa_0} \frac{mv^2}{2k_B T_0}{n_e^*}^{\frac{1}{\kappa_0+1}}
            }
    \right)
          \right] \notag \\
          =& \frac{\nabla n_e^*}{n_e^*} 
        \frac{1 + \frac{d}{2} \frac{1}{\kappa_0+1}}{
            1+\frac{1}{\kappa_0} \frac{mv^2}{2k_B T_0}{n_e^*}^{\frac{1}{\kappa_0+1}}
            }f_e,
            \label{ap:dfdx}
\end{align}
and 
\begin{equation}
   \pdv[]{f_e}{\vb{v}} = -\frac{\kappa_0+1+\frac{d}{2}}{\kappa_0}
        \frac{
                f_e {n_e^*}^{\frac{1}{\kappa_0+1}}
            }{
                1+\frac{1}{\kappa_0} \frac{mv^2}{2k_B T_0}{n_e^*}^{\frac{1}{\kappa_0+1}}
            }
        \frac{m\vb{v}}{k_BT_0}.
        \label{ap:dfdv}
\end{equation}
Substituting Eqs. \eqref{ap:dphi}, \eqref{ap:dfdx} and \eqref{ap:dfdv} into the left side of the Vlasov equation \eqref{eq:V}, we find all the terms vanish. 
Therefore, $f_e$ given by Eq. \eqref{eq:pdf} with $n_e$ from \eqref{eq:u} is a stationary solution of the Vlasov-Poisson system.
\section{The difference between the calculations from initial distribution and final simulated distribution}%
\label{ap:com}
\begin{table}[h]
   \centering
   \begin{tabular}{rllll}
   \hline
   \hline
                   & $E$~~~     & $S$~~~     & $\kappa_0$~~~ & $T_0$~~~ \\ 
   \hline
    Pure theoretical results~~~ & 1.41189~~~ & 5.41864~~~ & 0.147081~~~   & 2.83690~~~ \\
    Semi-theoretical results~~~   & 1.43744~~~ & 5.68090~~~ & 0.313970~~~   & 2.89604~~~ \\
   \hline
   \hline
   \end{tabular}
   \caption{Comparisons of parameters calculated from different ways for Case A.}
   \label{tab:com-para}
\end{table}
\begin{figure}[h]
   \centering
   \subfigure[]
   {
       \includegraphics[width=0.4\textwidth]{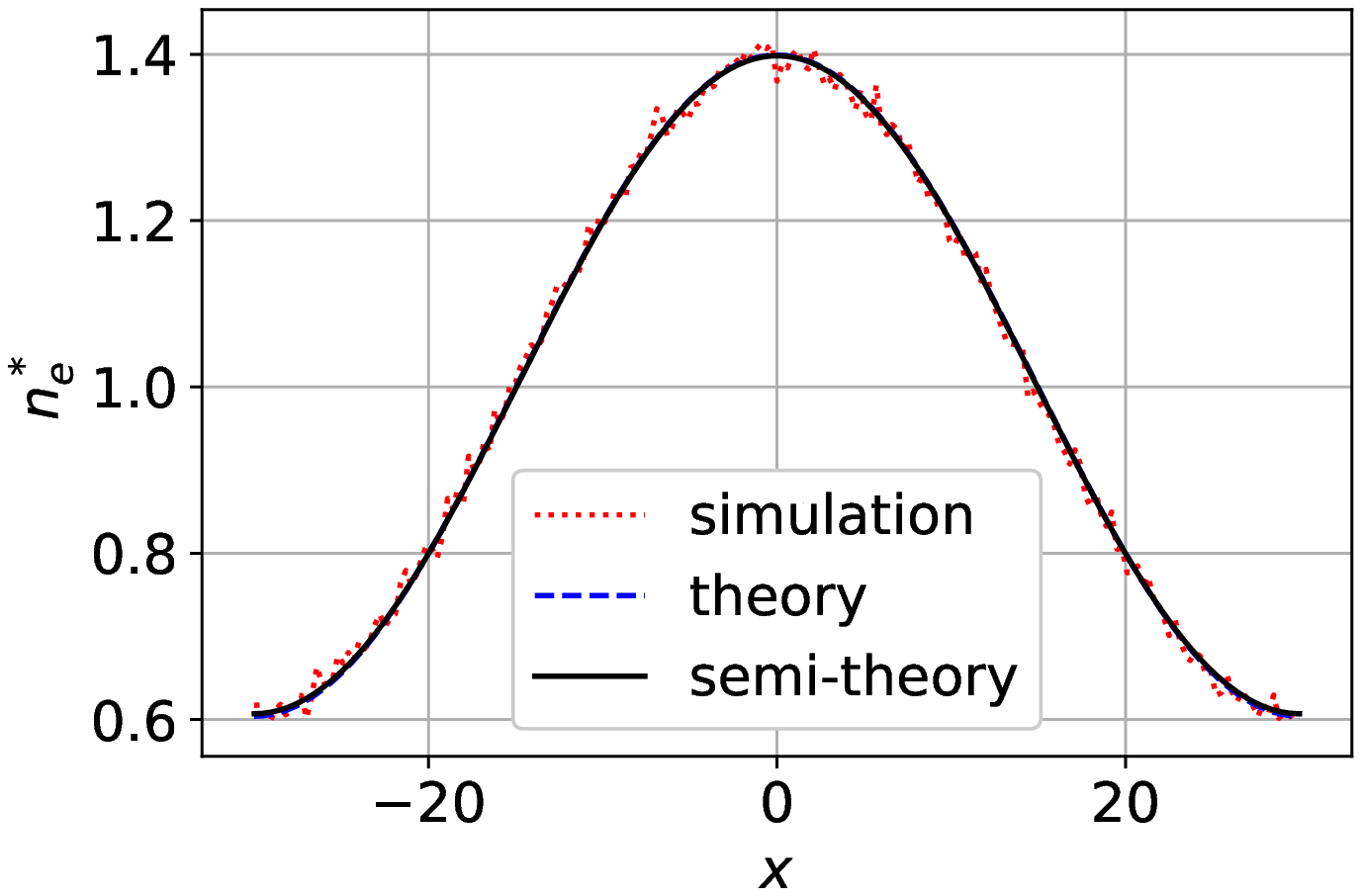}
       \label{fig:com-n}
   }
   \subfigure[]
   {
       \includegraphics[width=0.4\textwidth]{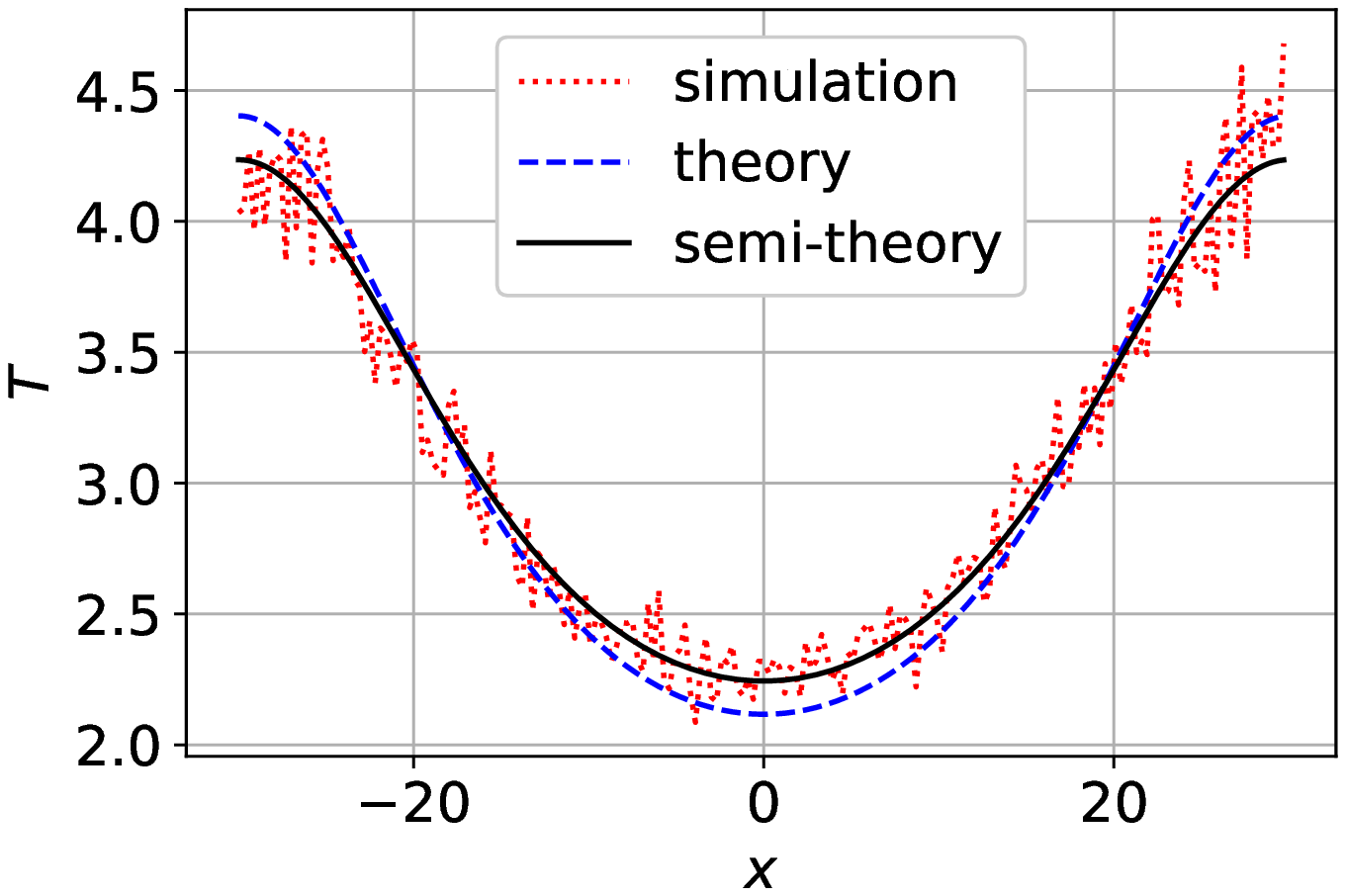}
       \label{fig:com-t}
   }
   \subfigure[]
   {
       \includegraphics[width=0.4\textwidth]{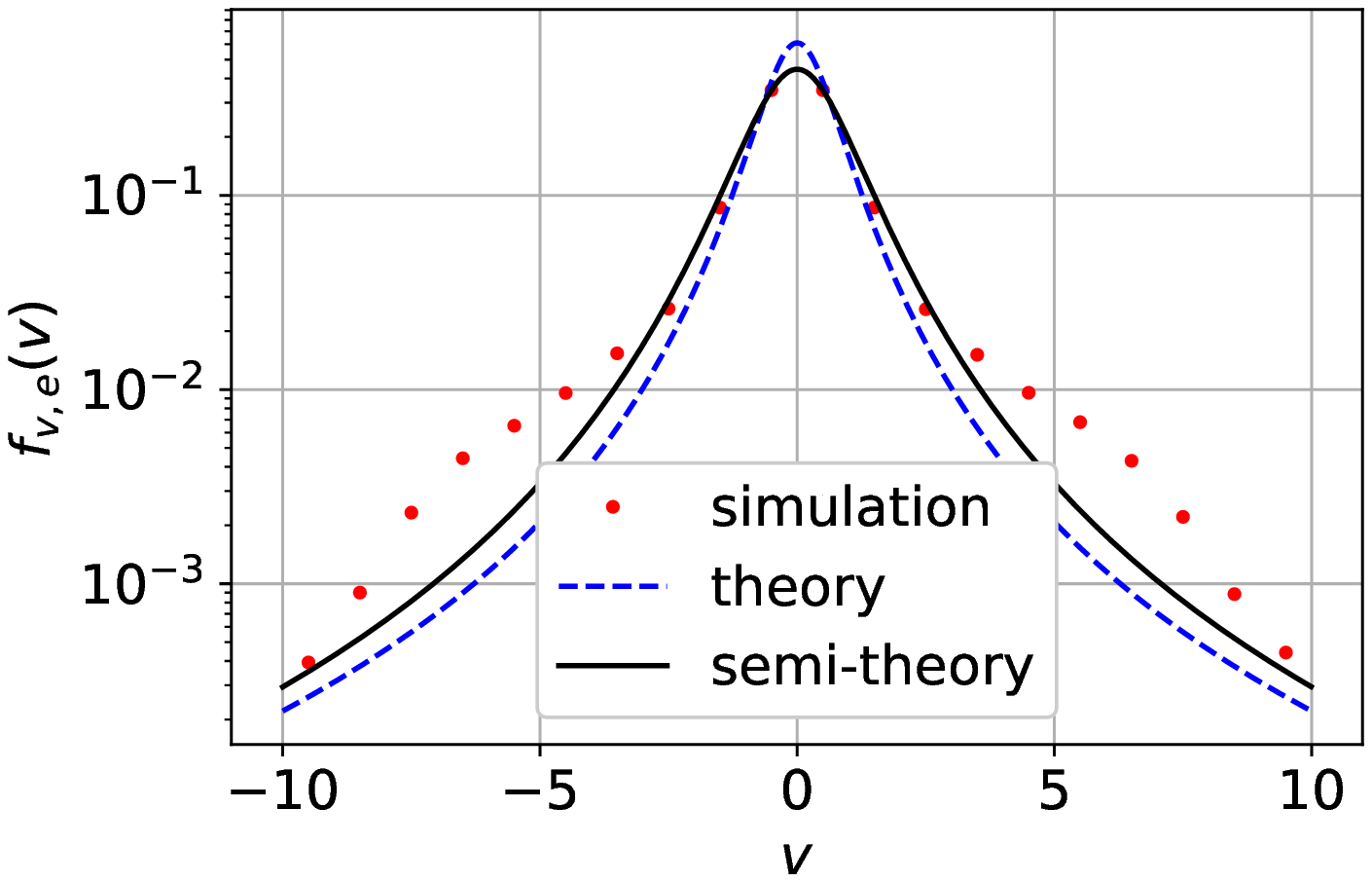}
       \label{fig:com-vpdf}
   }
   \caption{
       Comparisons of (a) the relative number density, (b) the temperature, and (c) the global velocity distribution.
       It shows that the theoretical solutions of number density are not sensitive for different values of energy and entropy. 
       But for the temperature distribution as well as the global velocity distribution, the theoretical results are obviously affected.
       These figures imply that semi-theoretical predictions are more accurate.
   }
   \label{fig:com}
\end{figure}
To calculate the parameters $\kappa_0$ and $T_0$, we need to obtain two constants of motion, which are chosen as the energy $E$ and the Boltzmann-Gibbs entropy $S$. 
There are two ways to calculate the energy and entropy. 
The first way is to calculate them from the theoretical initial distribution which is an isothermal Maxwellian distribution with inhomogeneous density.
The second way is to get them from the final distribution provided by the simulations.
In a real collisionless plasma, there is no difference between the first and second ways because both energy and entropy are conservations.
However, since the effects of self-heating in PIC simulations, the energy and entropy are slightly increasing during the time evolution, which means the values of them are different by using different ways of calculation.
The parameters $\kappa_0$ and $T_0$, as well as other theoretical results, are also different by employing different values of energy and entropy.
For the reason of distinction, we call them “pure theoretical predictions” for those results from the first way of calculation, and “semi-theoretical predictions” for those from the second way.
We here present the comparisons of these two kinds of theoretical predictions as shown in the following Table \ref{tab:com-para} and Fig. \ref{fig:com}.
The comparisons are made by only taking the parameters of Case A in the main body.
The results suggest that $T_0$ is influenced to a small extent due to a small difference in energy, but $\kappa_0$ is affected to a large extent because of a large difference in entropy. 

\bibliography{mylib}
\end{document}